\documentclass[apj]{emulateapj}
\usepackage{apjfonts}
\slugcomment{Accepted by the Astrophysical Journal}
\shortauthors{Tran et al.}
\shorttitle{\mipsmu~Sources in a Super Galaxy Group at $z=0.37$}

\begin{document}

\newcommand{\hi}{$h^{-1}$}
\newcommand{\kms}{~km~s$^{-1}$}
\newcommand{\mipsmu}{$24\mu$m}
\newcommand{\sfrir}{SFR$_{IR}\geq3M_{\odot}$~yr$^{-1}$}
\newcommand{\magcut}{$-20.5$}
\newcommand{\masscut}{$4\times10^{10}M_{\odot}$}

\title{A Spectroscopically Confirmed Excess of 
\mipsmu~Sources in a Super Galaxy Group at $z=0.37$:  Enhanced
Dusty Star Formation Relative to the Cluster and Field Environment}

\author{Kim-Vy H. Tran\altaffilmark{1,2}, Am\'elie
Saintonge\altaffilmark{1}, John Moustakas\altaffilmark{3,4}, Lei
Bai\altaffilmark{5,6}, Anthony  H. Gonzalez\altaffilmark{7}, Bradford
P. Holden\altaffilmark{8}, Dennis Zaritsky\altaffilmark{5}, \& Stefan
J. Kautsch\altaffilmark{7}}  

\altaffiltext{1}{Institute for Theoretical Physics, University of
  Z\"urich, CH-8057 Z\"urich, Switzerland}
\altaffiltext{2}{George P. and Cynthia W. Mitchell Institute for
Fundamental Physics and Astronomy, Department of Physics,  
Texas A\&M University, College Station, TX 77843; vy@physics.tamu.edu}
\altaffiltext{3}{Center for Cosmology and Particle Physics, 4
Washington Place, New York University, New York, NY 10003}
\altaffiltext{4}{Center for Astrophysics and Space Sciences,
University of California, San Diego, 9500 Gilman Drive, La Jolla,
California, 92093}  
\altaffiltext{5}{Steward Observatory, University of Arizona, 933 
North Cherry Avenue, Tucson, AZ 85721}
\altaffiltext{6}{Observatories of the Carnegie Institution of
Washington, 813, Santa Barbara Street, Pasadena, CA 91101}
\altaffiltext{7}{Department of Astronomy, University of Florida,
  Gainesville, FL 32611}
\altaffiltext{8}{UCO/Lick Observatories, University of California,
Santa Cruz, CA 95064}

\setcounter{footnote}{8}

\begin{abstract}

To trace how dust-obscured star formation varies with environment, we
compare the fraction of \mipsmu~sources in a super galaxy group to the
field and a rich galaxy cluster at $z\sim0.35$.  We draw on
multi-wavelength observations\footnote{Based on observations made with
1) The ESO Telescopes at Paranal Observatories under program IDs
072.A-0367, 076.B-0362, 078.B-0409; 2) the NASA/ESA Hubble Space
Telescope (GO-10499); STScI is operated by the association of
Universities for Research in Astronomy, Inc. under the NASA contract
NAS 5-26555; 3) the Spitzer Space Telescope, which is operated by the
Jet Propulsion Laboratory, California Institute of Technology under a
contract with NASA; support for this work was provided by NASA through
an award issued by JPL/Caltech (GO-20683); 4) the Chandra X-ray
Observatory Center, which is operated by the Smithsonian Astrophysical
Observatory for and on behalf of the National Aeronautics Space
Administration under contract NAS8-03060; and 5) the Magellan 6.5 m
telescope operated by OCIW.} that combine $Hubble$, $Chandra$, and
$Spitzer$ imaging with extensive optical spectroscopy ($>1800$
redshifts) to isolate galaxies in each environment and thus ensure a
uniform analysis.  We focus on the four galaxy groups
($\sigma_{1D}=303-580$\kms) in supergroup 1120-12 that will merge to
form a galaxy cluster comparable in mass to Coma.  We find that 1) the
fraction of supergroup galaxies with \sfrir~is four times higher than
in the cluster (32$\pm5$\% vs. 7$\pm2$\%); 2) the supergroup's
infrared luminosity function confirms that it has a higher density of
IR members compared to the cluster and includes bright IR sources
($\log(L_{IR})$[erg~s$^{-1}$]$>45$) not found in galaxy clusters at
$z\lesssim0.35$; and 3) there is a strong trend of decreasing
\mipsmu~fraction with increasing galaxy density, $i.e.$ an
infrared-density relation, not observed in the cluster.  These
dramatic differences are surprising because the early-type fraction in
the supergroup is already as high as in clusters, $i.e.$ the
timescales for morphological transformation cannot be strongly coupled
to when the star formation is completely quenched.  The supergroup has
a significant fraction ($\sim17$\%) of luminous, low-mass
($10.0<\log(M_{\ast})[M_{\odot}]<10.6$), \sfrir~members that are
outside the group cores ($R_{proj}\geq0.5$~Mpc); once their star
formation is quenched, most will evolve into faint red galaxies.  Our
analysis indicates that the supergroup's \mipsmu~population also
differs from that in the field: 1) despite the supergroup having twice
the fraction of E/S0s as the field, the fraction of \sfrir~galaxies is
comparable in both environments, and 2) the supergroup's IR luminosity
function has a higher $L_{IR}^{\ast}$ than that previously measured
for the field.

\end{abstract}


\notetoeditor{This is Paper II in a series where Paper I was
``Spitzer/MIPS 24 micron Observations of Galaxy Clusters: An Increasing
Fraction of Obscured Star-forming Members from z = 0.02 to z = 0.83''
by Saintonge, Tran, and Holden 2007, ApJ, DOI(10.1086/592730)}

%%%%%%%%%%%%%%%%%%%%%%%%%%%%%%%%%%%%%%%%%%%%%%%%%%%%%%%%%%%%%%%%%%%%
\section{Introduction}

Galaxies in the field environment span a wide range in morphology,
color, and ongoing star formation \citep[$e.g.$][]{marzke:98}.  In
contrast, the significantly more crowded environment of galaxy
clusters is dominated by passive, red, early-type galaxies that formed
the bulk of their stars at $z>2$
\citep{gregory:78,dressler:80,bower:92,vandokkum:98b}.  Using
spectroscopically defined samples, several studies show that galaxies
in clusters differ from their field counterparts even up to $z\sim1.4$
\citep[$e.g.$][]{holden:07,mei:09,lidman:08}.

However, we know that the galaxy populations in clusters have evolved
since at least $z\sim1$.  Observational examples of how the galaxy mix
in clusters evolves with increasing redshift include: the increasing
fraction of blue/star-forming members \citep{butcher:78,ellingson:01};
the increasing fraction of spectroscopically confirmed \mipsmu~sources
\citep[Paper I]{saintonge:08}; the increasing fraction of massive
post-starburst members \citep{tran:03b,poggianti:04}; the increasing
fraction of active galactic nuclei \citep{eastman:07,kocevski:09}; the
increasing fraction of star-forming galaxies with increasing galaxy
density at $z\sim1$, a reversal of what is observed at $z\sim0$
\citep{elbaz:07,cooper:08}; the decreasing fraction of S0 galaxies
\citep{postman:05,moran:07}; and the decreasing fraction of faint red
galaxies
\citep[$z>0.5$;][cf. \citealt{crawford:09}]{tanaka:07,stott:07,delucia:07a}.

The problem is that we have yet to identify clearly the physical
mechanisms responsible for the dramatically different galaxy
populations in clusters versus the field, nor the timescales needed
for these mechanisms to operate.  Although a plethora of physical
processes have been invoked to quench star formation and transform
galaxies into spheroidal systems, $e.g.$ starvation \citep{bekki:02},
ram-pressure stripping \citep{abadi:99}, and galaxy harassment
\citep{moore:98}, stringent observational tests of when star formation
is quenched and whether quenching is coupled to morphological
transformation are needed to assess the relative importance of the
physics at work.  Simulations are sufficiently advanced that new
insight can be obtained by, $e.g.$ comparing star formation and
gas-loss rates as a function of local density to the observations
\citep{tonnesen:07}.

Also, instead of focusing on massive clusters, the key to
understanding the interplay between galaxy evolution and environment
is to study galaxy groups because: 1) most galaxies in the local
universe are in groups \citep[$e.g.$][]{geller:83}; and 2)
hierarchical structure formation predicts that galaxy clusters
assemble from the merger and accretion of smaller structures, $e.g.$
groups \citep{peebles:70}.  Simulations show that the physical
mechanisms normally associated with galaxy clusters are also effective
in groups \citep{hester:06,kawata:08,mccarthy:08}, but that they
operate in groups at lower redshifts \citep{romeo:08}.

In fact, the galaxy groups in the local universe do have more in
common with galaxy clusters than with the field population, $i.e.$
higher early-type fractions and lower mean star formation rates than
the field
\citep{zabludoff:98a,hashimoto:98,tran:01,blanton:07b,rasmussen:08}.
With the advent of large spectroscopic studies such as CNOC
\citep{yee:96} and SDSS \citep{sdss1:03}, the galaxy populations in
groups can now be studied for statistically large samples
\citep[$e.g.$][]{yang:07}, and at intermediate redshifts
\citep[$e.g.$][]{poggianti:08,gal:08,knobel:09}.  Although still
nascent, spectroscopic studies of galaxy groups at $z>0.3$ find that
the groups already have high early-type fractions
\citep{jeltema:07,wilman:08}.  However, whether star formation in
$z>0.3$ groups is enhanced or simply quenched relative to the field is
debated \citep{poggianti:09,balogh:09}.

The question then is whether the evolution of galaxies in clusters is
driven primarily on group or on cluster scales.  Our discovery of a
supergroup of galaxies at $z=0.37$ allows us to uniquely answer this
question.  The supergroup (hereafter SG1120) is composed of multiple
galaxy groups that we have shown will merge into a cluster comparable
in mass to Coma by $z\sim0$ \citep{gonzalez:05}, unlike the majority
of clusters studied at $z>0.3$ that are too massive to be Coma
progenitors.  Because we know the galaxies in SG1120 will evolve into
a cluster population, we can test whether the group galaxies are
already like those in clusters.  First results from our
multi-wavelength study of SG1120 show that the group galaxies are in
transition: SG1120 has a high fraction of early-type members
\citep{kautsch:08}, yet several of the most massive group galaxies are
growing by dissipationless merging at $z<0.4$ \citep{tran:08}.

Here we focus on the dust-obscured star formation in the supergroup as
measured with MIPS \citep{rieke:04} \mipsmu~observations.  Studies
find a surprising number of mid-infrared sources at cluster and group
densities at $z>0.3$ \citep{elbaz:07,bai:07,koyama:08,dressler:09},
but this may be due to the general increase in the fraction of
mid-infrared galaxies with redshift \citep{lefloch:05}.  To determine
if there is an excess of IR sources in the galaxy groups making up
SG1120, we compare the \mipsmu~members in SG1120 to their counterparts
in both the field and cluster environment at the same redshift.

For the cluster environment, we use CL~1358+62, a massive, dynamically
evolved, X-ray luminous galaxy cluster at $z=0.328$ with a
line-of-sight velocity dispersion of
$1027^{+51}_{-45}$\kms~\citep[hereafter F98]{fisher:98}.  Most of the
232 spectroscopically confirmed members are passive, early-type
galaxies \citep{vandokkum:98a}.  In addition to the spectroscopy, we
have HST/WFPC2 mosaics taken in F606W and F814W covering $\sim50\Box'$,
and MIPS \mipsmu~imaging.

The field sample is drawn from extensive observations of two higher
redshift clusters: MS~2053--04 \citep[$z=0.59$;][]{tran:05a} and
MS~1054--03 \citep[$z=0.83$;][]{tran:07}.  In the combined area of
approximately $50\Box'$, we have measured spectroscopic redshifts for
nearly 300 field galaxies at $0.09<z<1.36$.  These fields were also
imaged with HST/WFPC2 in F606W and F814W, and with MIPS at \mipsmu.  The
depth and uniformity of our spectroscopic and photometric observations
in these fields makes for a unique dataset that enables robust
comparison across environment at $z\sim0.35$.

Our study of how \mipsmu~galaxies vary across environment is Paper II
in our SMIRCS ($Spitzer$/MIPS Infra-Red Cluster Survey) series and
complements Paper I \citep{saintonge:08} where we explored how the
\mipsmu~fraction increases with increasing redshift in massive galaxy
clusters at ($0<z<0.83$; see also Finn et al., in prep).  Throughout
the paper, we use $H_0=70$~km~s$^{-1}$~Mpc$^{-1}$, $\Omega_M=0.3$, and
$\Omega_{\Lambda}=0.7$; at $z=0.37$, this corresponds to a scale of
5.12 kpc per arcsec and a lookback time of 4 Gyr.  All restframe
magnitudes are in the Vega system.

%%%%%%%%%%%%%%%%%%%%%%%%%%%%%%%%%%%%%%%%%%%%%%%%%%%%%%%%%%%%%%%%%%%%
\section{Observations}

At these redshifts ($z\sim0.35$), our large number of
spectroscopically confirmed group (174) and cluster members (232)
combined with multi-wavelength imaging that includes MIPS observations
is unique among existing surveys.  Unlike many spectroscopic cluster
surveys at intermediate redshifts, we do not select by optical color
which can bias a sample towards members that are already on the red
sequence, $e.g.$ against blue, star-forming members.  The depth of our
redshift surveys in the cluster fields also enables us to identify a
sample of field galaxies ($0.25\leq z\leq0.45$; $\bar{z}=0.35$) that
have been observed and analyzed in the same manner as the group and
cluster galaxies.  The uniformity of our observations allows us to
compare directly galaxy populations across a range of environments at
$z\sim0.35$.

\subsection{Optical Imaging}

\subsubsection{Supergroup 1120 ($z\sim0.37$)}

The four X-ray luminous galaxy groups in the supergroup 1120-12
(hereafter SG1120) extend across an approximately $8'\times12'$ region
(Fig.~\ref{sg1120-xy}).  Optical photometry of the group galaxies is
measured from VLT/VIMOS \citep{lefevre:03} mosaics in $BVR$
($18'\times20'$; (PSF)$_R\sim0.5''$), Magellan/LDSS3 mosaics in $g'r'$
($12'\times20'$; (PSF)$_{r'}\sim1.0''$), and a 10 pointing mosaic
taken with HST/ACS in F814W ($11'\times18'$; $0.05''$/pixel).
Near-infrared imaging was also obtained with
KPNO/FLAMINGOS\footnote[10]{FLAMINGOS was designed and constructed by the
IR instrumentation group (PI: R. Elston) at the University of Florida,
Department of Astronomy, with support from NSF grant AST97-31180 and
Kitt Peak National Observatory.} and provides a $K_s$ mosaic
($16'\times19'$; PSF$\sim1.2"$).  The wide-field mosaics are generated
with {\it scamp} and {\it swarp}\footnote[11]{http://astromatic.iap.fr}
\citep{bertin:02,bertin:06} which corrects the astrometry across the
wide field and stitches the pointings together.

Line-matched photometric catalogs were generated using the VIMOS $R$
mosaic as the master detection image \citep[SExtractor
v2.5.0;][]{bertin:96}. While several close galaxy pairs
(separation$<1''$) are considered single objects in the $R$ catalog,
this is appropriate for our analysis given that the same close pairs
are also blended sources in the MIPS catalog (see \S\ref{mipstext}).
We use {\it k-correct} v4.1 \citep{blanton:07a} to determine
rest-frame absolute magnitudes (Vega) and K-corrections.  As input, we
use the MAG$_{-}$AUTO photometry from the $g'BVr'R$ imaging and
assumed minimum photometric uncertainties in each bandpass of 0.05
mag. The photometry has been corrected for foreground Galactic
extinction using the \citet{schlegel:98} dust maps and the
\citet{odonnell:94} Milky Way extinction curve, assuming $R_V=3.1$.

For consistency and to thus ensure that our comparisons are robust
across the supergroup, cluster, and field samples, we calculate
stellar masses in the same manner by following the prescription in
\citet{bell:03b}.  Here mass-to-light ratios $(M_{\ast}/L)_B$ are
calculated from $(B-V)$ colors using:

\begin{equation}
\log(M_{\ast}/L)_B=1.737(B-V)-0.942
\end{equation}

\noindent and a diet Salpeter initial mass function (IMF) is assumed.
We use the diet Salpeter IMF defined in \citet{bell:01} as having
$x=0$ below $0.6M_{\odot}$ and so the stellar mass using a diet
Salpeter IMF is 70\% of that for a regular Salpeter IMF
\citep{salpeter:55}.  Using an absolute magnitude for the sun of
$M_B=5.45$\footnote[12]{http://www.ucolick.org/$\sim$cnaw/sun.html}, a
galaxy with $M_B=-19.5$ and $(B-V)=1$ has a stellar mass of
$\log(M_{\ast})[M_{\odot}]=10.8$.

\subsubsection{CL~1358+62 ($z=0.328$)}

The galaxy populations in CL~1358+62 ($z=0.328$; F98) have been
studied extensively using optical imaging and spectroscopy.  For the
cluster galaxies, we use the optical photometry measured by
\citet[hereafter H07]{holden:07} from the HST/WFPC2 mosaics (total
area of $\sim50\Box'$).  To summarize, a S\'ersic profile ($1\leq
n\leq4$) was fit to the surface brightness distribution in the
HST/WFPC2 imaging of each spectroscopically confirmed member; over
85\% of the cluster members have $n\geq2$.  Galaxy colors were
determined from fluxes measured within a half-light radius; the
half-light radii were determined using the F814W imaging.  Note that
at $z\sim0.33$, the redshifted $B$ and $V$ filters are well-matched to
F606W and F814W.  As in SG1120, we convert the observed fluxes
(corrected for Galactic extinction) to rest-frame $BV$ magnitudes
using $k-correct$, and estimate stellar masses using Eq. 1.  For more
details about the photometry and testing the robustness of the stellar
mass determination for the cluster galaxies, we direct the reader to
the extensive dicussion in H07.

\subsubsection{Field Galaxies ($\bar{z}=0.35$)}

Our field sample is drawn from a larger program that focused on
galaxies in X-ray luminous clusters at intermediate redshifts.  To
select field galaxies in the same redshift range as SG1120 and CL1358,
we use observations of galaxy clusters MS~2053--04
\citep[$z=0.59$;][]{tran:05a} and MS1054--03
\citep[$z=0.83$;][]{tran:07}.  Both galaxy clusters were imaged by
HST/WFPC2 in the F606W and F814W filters; each image mosaic is
composed of six overlapping pointings and each mosaic covers an area
of $\sim25\Box'$.  The image reduction and photometry are detailed for
MS2053 and MS1054 in \citet{hoekstra:02} and \citet{vandokkum:00},
respectively.  

Photometric catalogs were generated using SExtractor
\citep[see][]{tran:04a} and we used $k-correct$ to convert observed
fluxes (measured with MAG\_AUTO and corrected for Galactic extinction)
to rest-frame $BV$ magnitudes.  As in the cases of the supergroup and
cluster galaxies, stellar masses for the field galaxies are estimated
using Eq. 1.

\subsection{Optical Spectroscopy}\label{redshifts}

\subsubsection{Supergroup 1120 ($0.34<z<0.38$)}

The spectroscopic survey of the SG1120 field was completed using
VLT/VIMOS (in 2003), Magellan/LDSS3 (in 2006), and VLT/FORS2
\citep[in 2007;][]{appenzeller:98}.  The medium resolution spectroscopy
corresponds to 2.5\AA~pix$^{-1}$ (VIMOS), 0.7\AA~pix$^{-1}$ (LDSS3),
and 1.65\AA~pix$^{-1}$ (FORS2).  Targets for the VIMOS masks were
selected using $R\leq22.5$ mag, and targets for the later runs
selected using $K_s\leq20$ mag.  A total of 16 slit-masks were
observed at varying position angles, thus our spectroscopic
completeness is not affected by slit collisions.

Spectra from all of the observing runs were reduced using a
combination of IRAF\footnote[13]{IRAF is distributed by the National
Optical Astronomy Observatories.} routines and custom software
provided by D. Kelson \citep{kelson:00b}; see \citet{tran:05a} for
further details on the spectral reductions.  Redshifts were determined
using IRAF cross-correlation routines, and each assigned redshift was
visually compared to the 1D spectrum.  Each redshift was then given a
quality flag where $Q=3, 2, \&~1$ corresponded to definite, probable,
and maybe (single emission line).  The spectral range for most of the
supergroup members covers [OII]$\lambda3727$ to [OIII]$\lambda5007$.

\begin{figure}
\epsscale{1.0}
\plotone{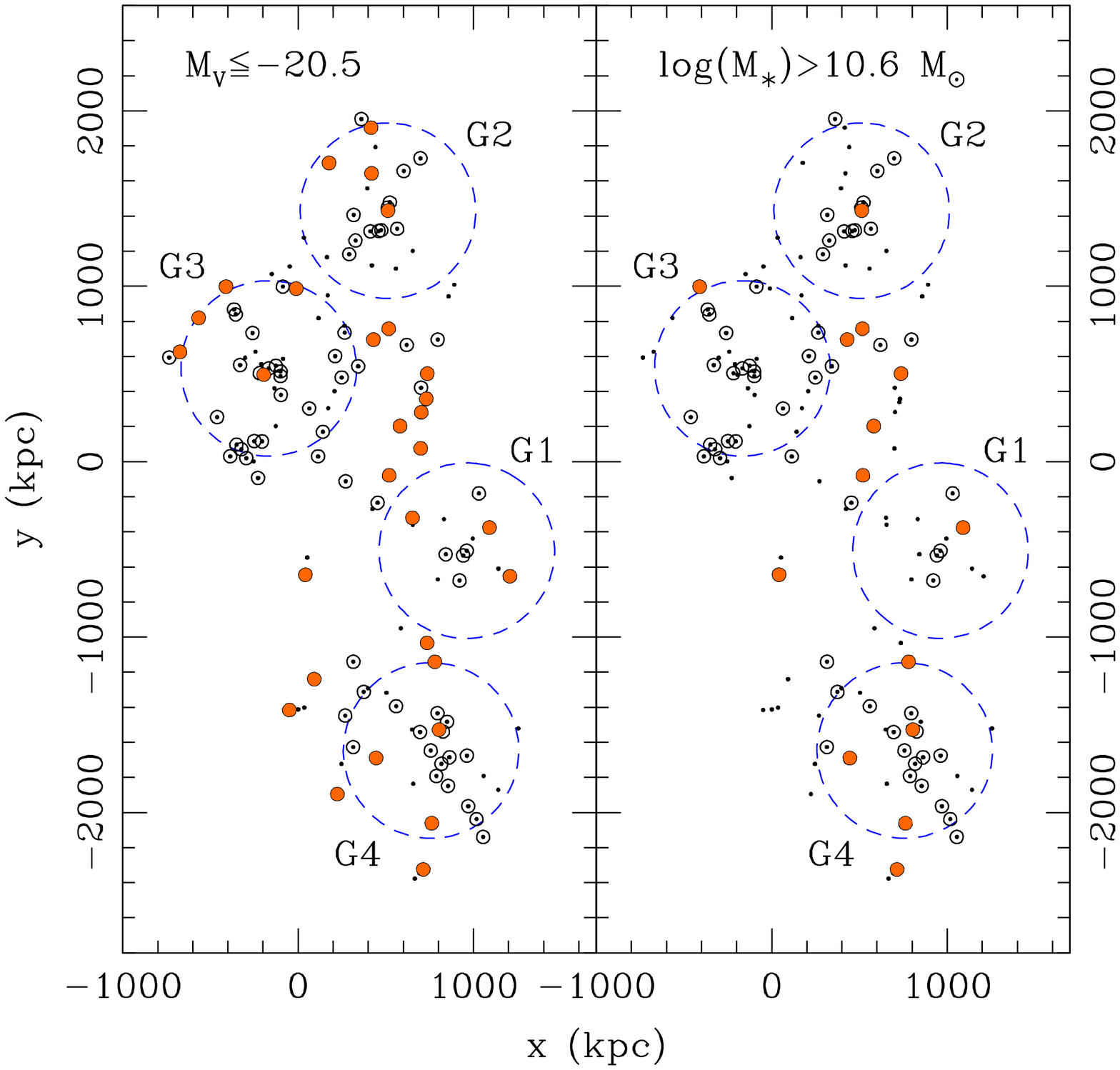}
\caption{Spatial distribution of the spectroscopically confirmed
supergroup galaxies that fall on the HST/ACS mosaic (small dots;
$0.34\leq z\leq0.38$; 143 of 174 total members); north is up and 
east to the left.  {\it Left:} Open circles denote all group galaxies brighter
than $M_V=$\magcut, and large filled circles denote members brighter
than $M_V=$\magcut~with \sfrir.  The large dashed circles ($R=0.5$\hi
Mpc) are centered on the brightest group galaxies listed in
Table~\ref{tab:groups}; these positions are well-matched to the
extended X-ray emission detected in each group. {\it Right:} The same
but for the mass-selected ($M_{\ast}>4\times10^{10}M_{\odot}$)
supergroup members.  Note the number of luminous, \mipsmu~members
outside the group cores that are low-mass systems ($10.0<
\log(M_{\ast})[M_{\odot}]<10.6$). 
\label{sg1120-xy}}
\end{figure}

\begin{figure}
\epsscale{1.0}
\plotone{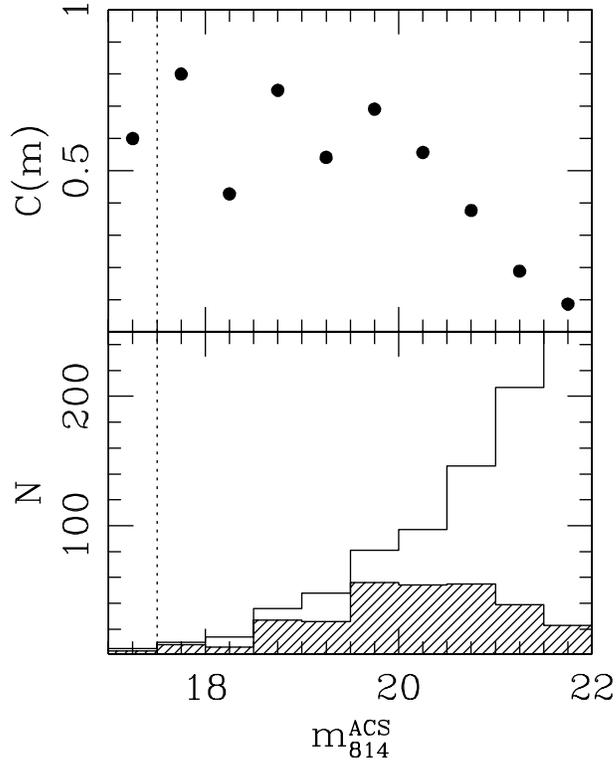} 
% in sg1120/john/Line-matched-catalogs/V0-06.06.2008/members-mags.sm
\caption{{\it Top:} For the SG1120 region, the spectroscopic
completeness in the HST/ACS mosaic shown as a function of F814W
magnitude; the dotted vertical line denotes the magnitude of the
brightest group galaxy ($m_{814}^{ACS}=17.5$).  {\it Bottom:} Histograms
showing the total number of galaxies in the HST/ACS field (open), and
those with redshifts (shaded).  Variations in $C(m)$ at magnitudes
brighter than 18.5 are due to small numbers ($<5$) in each magnitude
bin.  The spectroscopic survey is $>50$\% complete at
$m_{814}^{ACS}<20.5$; our adopted magnitude limit of
$M_V=$\magcut~mag corresponds approximately to $m_{814}^{ACS}=20.4$.
\label{Cmhist}}
\end{figure}

The spectroscopic completeness in the HST/ACS footprint is shown in
Fig.~\ref{Cmhist}.  Due to the supergroup's elongated structure (see
Fig.~\ref{sg1120-xy}), spectra for a few of the bright
($m_{814}^{ACS}<18.5$) galaxies have not been obtained; however, these
are foreground galaxies.  The brightest group galaxy has
$m_{814}^{ACS}=17.5$ mag, and the survey remains $>50$\% complete to
$m_{814}^{ACS}=20.5$ mag.  For red supergroup members, the adopted
magnitude limit used in our analysis of $M_V=$\magcut~mag (see
\S\ref{text:magcut}) corresponds to $m_{814}^{ACS}=20.4$ mag.

\begin{figure}
\epsscale{1.0}
\plotone{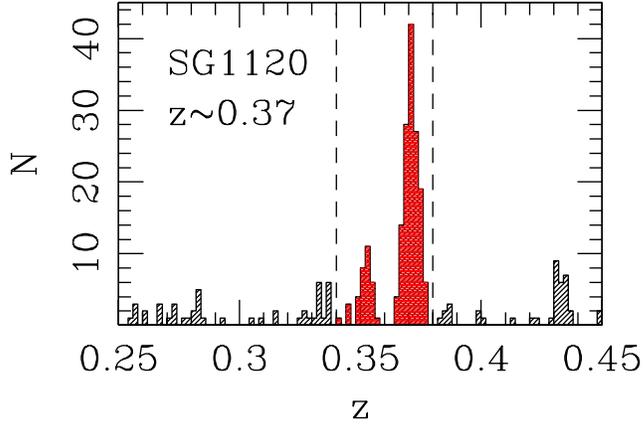}
% in /sg1120/Spectra/Spectra-info/plots.sm
\caption{Redshift distribution of galaxies in the supergroup (SG1120)
field; the redshift range for group members ($0.34\leq z\leq0.38$) is
denoted by the vertical dashed lines.  The brightest group galaxies
lie at $0.354\leq z\leq0.372$.
\label{sg1120-zhist}}
\end{figure}

In the larger $20'\times20'$ region centered on the HST/ACS mosaic, we
have obtained spectra of 603 unique objects.  Guided by breaks in the
redshift distribution, we define group members to be at $0.34\leq
z\leq0.38$ (Fig.~\ref{sg1120-zhist}).  Considering only galaxies with
redshift quality flag of $Q=3$ gives 174 supergroup members.  Four of
the five X-ray luminous regions correspond to galaxy groups at
$0.35<z<0.37$ while the fifth is a galaxy cluster at $z=0.48$
\citep{gonzalez:05}.  The coordinates, mean redshifts, and
line-of-sight velocity dispersions of the individual groups are listed
in Table~\ref{tab:groups}, and Fig.~\ref{sg1120-xy} shows the spatial
distribution of members on the HST/ACS mosaic.

\subsubsection{CL~1358+62 ($0.315<z<0.342$)}\label{text:magcut}

A complete description of the spectroscopic survey in CL1358 including
target selection, spectral reduction, wavelength calibration, sky
subtraction, etc., is presented by F98.  To summarize, WHT and MMT
spectroscopy targeted objects with $R\leq21$ mag over a $10'\times11'$
region; at this magnitude limit, the spectroscopic survey is $>80$\%
complete and not dependent on color (see F98, Fig. 2).  The magnitude
limit corresponds approximately to $M_V=$\magcut, and we use this
limit for our luminosity-selected samples.  For reference, the Coma
cluster has $M_V^{\ast}=-20.6$\footnote[14]{Here we use
$m_{V}^{\ast}=14.5$ \citep{abell:77} and a distance modulus of
$(m-M)=35.11$ \citep{baum:97} for the Coma cluster.}

From nearly 400 redshifts, cluster membership was confirmed for 232
galaxies; in our analysis, we consider only the 171 members that fall
on the HST/WFPC2 mosaic (total area of $\sim50\Box'$) that were
studied by H07.

\subsubsection{Field Galaxies ($0.25\leq z\leq0.45$)}

As part of our program on galaxy clusters at intermediate redshift, we
also obtained redshifts for a large sample of field galaxies.
Spectroscopic targets were selected using a magnitude cut of
$m_{814}^{WFPC2}\leq23$ and $m_{814}^{WFPC2}\leq23.5$ in the MS2053
and MS1054 fields respectively.  These magnitude-limited spectroscopic
surveys were completed with Keck/LRIS \citep{oke:95} and resulted in a
total of over 800 redshifts in the two fields; excluding the cluster
members and considering only redshifts with $Q=3$ provides 295 field
galaxies at $0.09<z<1.36$.  Further observational details for each
field are in \citet{tran:04a}.  Notably, the spectroscopic
completeness in both cluster fields is $>80$\% at $m_{814}^{WFPC2}<21$
\citep[see Figs. 2 in][]{tran:05a,tran:07}.

\begin{figure}
\epsscale{1.0}
\plotone{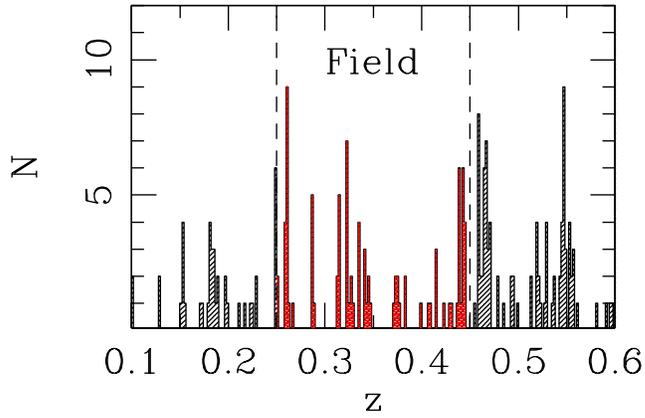}
% in /mips-members/SG1120-mips/plots.sm
\caption{The redshift distribution of the field galaxies at
$0.1<z<0.6$.  These field galaxies are drawn from extensive
spectroscopic 
surveys of two galaxy clusters, both of which are at $z>0.55$.  The
redshift bins are $\Delta z=0.002$ and are $5-10$ times larger than
the typical error in the redshift; any apparent overdensities
disappear with smaller bin sizes.  The vertical dashed lines denote
the redshift range of our comparison field sample ($0.25\leq z\leq0.45$).
\label{field_hist}}
\end{figure}

To ensure that we are observing the field galaxies at the same epoch
as the group and cluster galaxies, we use only the field galaxies at
$0.25\leq z\leq0.45$ ($\bar{z}=0.35$; Fig.~\ref{field_hist}).  In this
redshift range, we have 87 field galaxies; applying a magnitude
($M_V<$\magcut) or mass ($M_{\ast}\geq$\masscut) selection decreases
the field sample to 28 and 21 galaxies respectively (see
Table~\ref{tab:fractions}).  Note that both galaxy clusters MS2053
\citep[$z=0.59$;][]{tran:05a} and MS1054 \citep[$z=0.83$;][]{tran:07}
are at higher redshift, thus our field sample is not contaminated with
cluster galaxies.

\subsection{Hubble Type}

We have visually classified Hubble types that were assigned using HST
imaging for $>90$\% (371/401) of the spectroscopically-defined galaxy
sample across all three environments; at $M_V<$\magcut, an even higher
fraction (97\%; 225/232) of the galaxies are classified.  The high
resolution HST imaging allows us to easily separate bulge
vs. disk-dominated galaxies, and even to distinguish between
elliptical and S0s \citep{postman:05}.  We use a simplified Hubble
scheme where T-types are assigned to elliptical ($-5\leq T\leq-3$), S0
($-2\leq T\leq0$), spiral+irregular ($1\leq T\leq10$), and merging
($T=99$) galaxies.

% Field: 68/87 typed, 26/29 bright
% SG: 142/143 typed, 98/98 bright
% CL1358: 161/171 typed, 101/105 bright

In SG1120, we use the T-types assigned by \citet{kautsch:08} to 142 of
the 143 group galaxies that fall on the HST/ACS mosaic.  The galaxy
groups in SG1120 have velocity dispersions that are significantly
lower than in massive clusters such as CL1358
($303-580$\kms~vs. 1027\kms; see Table~\ref{tab:groups}), yet the
groups are already dominated by early-type members: SG1120's
early-type fraction of $\sim70$\% is already comparable to that of
galaxy clusters at intermediate redshifts \citep{kautsch:08}.

For the cluster (CL1358) and field galaxies, we have T-types assigned
by D. Fabricant, M. Franx, and P. van Dokkum using HST/WFPC2 imaging
\citep{fabricant:00}.  This team classified all galaxies in the
cluster fields brighter than $m_{814}^{WFPC2}=22$; these classifications have
been published in vD98, \citet{vandokkum:00}, and \citet{tran:05a}.
From this database, 161 of the 171 CL1358 galaxies in H07's sample and
67 of the 87 field galaxies ($0.25\leq z\leq0.45$) have visual
classifications.

\subsection{MIPS \mipsmu~Imaging}\label{mipstext}

Deep wide-field \mipsmu~imaging of all the fields presented in our
study was taken with MIPS \citep{rieke:04}.  We briefly summarize here
the procedure for retrieving, reducing, and analyzing the MIPS
imaging; further details are in Paper I.  For SG1120, we retrieved the
MIPS \mipsmu~data sets from the $Spitzer$ archive and corrected the
individual frames with the scan mirror position-dependent flats before
combining the frames with the MOPEX software to a pixel size of
$1.2''$.  The integration time per pixel was 1200 seconds and the
background level 35.5 MJy/sr.

With the large SG1120 MIPS mosaic ($22'\times56'$), we were able to
determine a good point spread function (PSF) and thus measure
\mipsmu~fluxes via profile fitting.  As a check, we compared the
fluxes measured via profile fitting to aperture fluxes and found the
values to be consistent; for the latter, we used an aperture diameter
of $2''$ as a compromise between maximizing the flux and minimizing
contamination from close neighbors, and applied corrections based on
fluxes derived from modeled PSFs (see Paper I).  We matched the
centroid position of the MIPS sources to the master $R$-band catalog.
We estimated the completeness of the SG1120 \mipsmu~catalog by adding
50 sources modeled on the empirical PSF to the mosaic and repeating
this process 20 times.

To convert the \mipsmu~fluxes to star formation rates, we determined
the total infrared luminosity ($F_{8-1000\mu {\rm m}}$) for each
galaxy using a family of infrared spectral energy distributions (SEDs)
from \citet{dale:02}. Using the range of SEDs that are representative
of galaxies in the Spitzer Infrared Nearby Galaxies Survey
\citep{dale:07}, we adopt the median conversion factor from
$F_{24\mu{\rm m}}$ to $F_{8-1000\mu{\rm m}}$ at $z\sim0.37$ where the
SEDs give essentially the same values and the error due to the adopted
conversion factor is only $\sim10-20$\%.  Combining this conversion
with the completeness simulations, we estimate that the SG1120
\mipsmu~catalog is 80\% complete to $\log(L_{IR})$[erg~s$^{-1}$]=43.8;
this corresponds to a \mipsmu~flux of approximately 105$\mu$Jy and an
IR star formation rate of \sfrir~\citep{rieke:09}.

For the smaller cluster fields, we followed essentially the same
procedure except that we used APEX to measure fluxes within a $2''$
diameter aperture and corrected the measured fluxes using the PSF from
the SG1120 mosaic.  The total integration times and background levels
in these mosaics vary, but the \mipsmu~imaging is essentially
confusion-limited in these fields (see Paper I).  We estimated the
completeness of the \mipsmu~catalogs by adding 30 sources into each
mosaic and repeated the process 20 times for each mosaic.  The \mipsmu~
catalogs are deeper than in the SG1120 field, $e.g.$ the CL1358
\mipsmu~catalog is 80\% complete to $\log(L_{IR})$[erg~s$^{-1}$]=43.5.

%%%%%%%%%%%%%%%%%%%%%%%%%%%%%%%%%%%%%%%%%%%%%%%%%%%%%%%%%%%%%%%%%%%%
\section{Results}

In the following analysis, we consider only the 143 supergroup
galaxies in SG1120 that fall on the HST/ACS mosaic, the 171 cluster
galaxies in CL1358 with photometry measured by H07 from the HST/WFPC2
mosaics, and the 87 field galaxies at $0.25\leq z\leq0.45$, all of
which also have HST/WFPC2 imaging.  We are thus assured of a uniformly
selected sample and can directly compare results across the three
environments.

Our \mipsmu~imaging identifies all galaxies with obscured
star-formation rates of $3M_{\odot}$~yr$^{-1}$ or greater, regardless
of galaxy mass.  However, actively star-forming galaxies tend to have
lower mass-to-light ratios than galaxies dominated by older stars,
$e.g.$ galaxies on the red sequence, thus an optical
luminosity-selected sample is likely to differ from a mass-selected
sample.  For this reason, we use both luminosity and mass-selected
samples in our analysis to check the robustness of our results.  Note
that in fitting the infrared luminosity functions, we select based on
total IR luminosity as determined with the \mipsmu~fluxes.

\subsection{Fraction of \mipsmu~Sources}\label{mipsfrac}

\begin{figure}
\epsscale{1.0}
\plotone{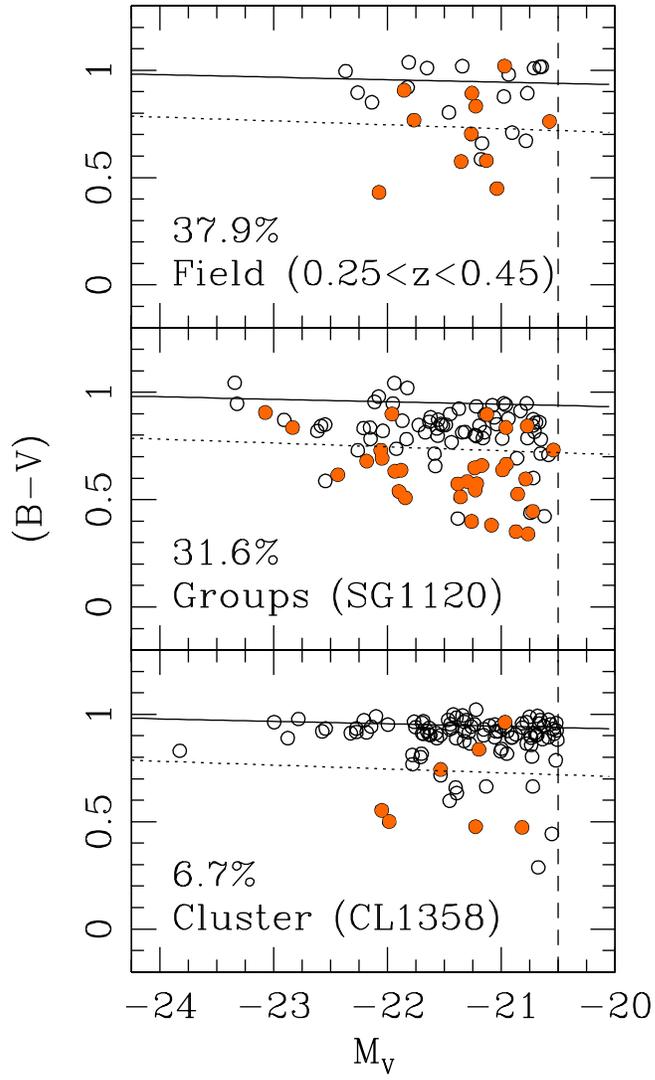}
\caption{Rest-frame color-magnitude (CM) diagram of galaxies brighter
than $M_V=$\magcut~(dashed vertical line) in the field (top), galaxy groups
(middle), and galaxy cluster (bottom). The galaxies with \sfrir~are
shown as filled circles; the corresponding fractions are listed in
each panel.  The slope of the CM relation (solid line) is from vD98
and normalized to the red sequence in CL1358; the same CM relation is shown
in each panel.  The \mipsmu~fraction in the galaxy groups is nearly as
high as in the field and is four times higher than in the cluster.
\label{cmd}}
\end{figure}

We first apply a luminosity limit set by the spectroscopic
completeness in the CL1358 field (see \S\ref{redshifts}) and consider
only galaxies with absolute $V$-band (Vega) magnitude brighter than
\magcut; due to the mixed galaxy population in our $z\sim0.35$
samples, we do not correct for passive evolution. The color-magnitude
(CM) diagram for the galaxies in all three environments is shown in
Fig.~\ref{cmd}.  The slope of the CM relation each panel is from vD98
who measured the CM relation in CL1358 using the early-type members;
the CM relation is normalized to the red sequence in CL1358.  The
color deviation from the CM relation is denoted as $\Delta(B-V)$ where
blue galaxies are classically defined has having $\Delta(B-V)<-0.2$
\citep{butcher:78}.

\begin{figure}
\epsscale{1.0}
\plotone{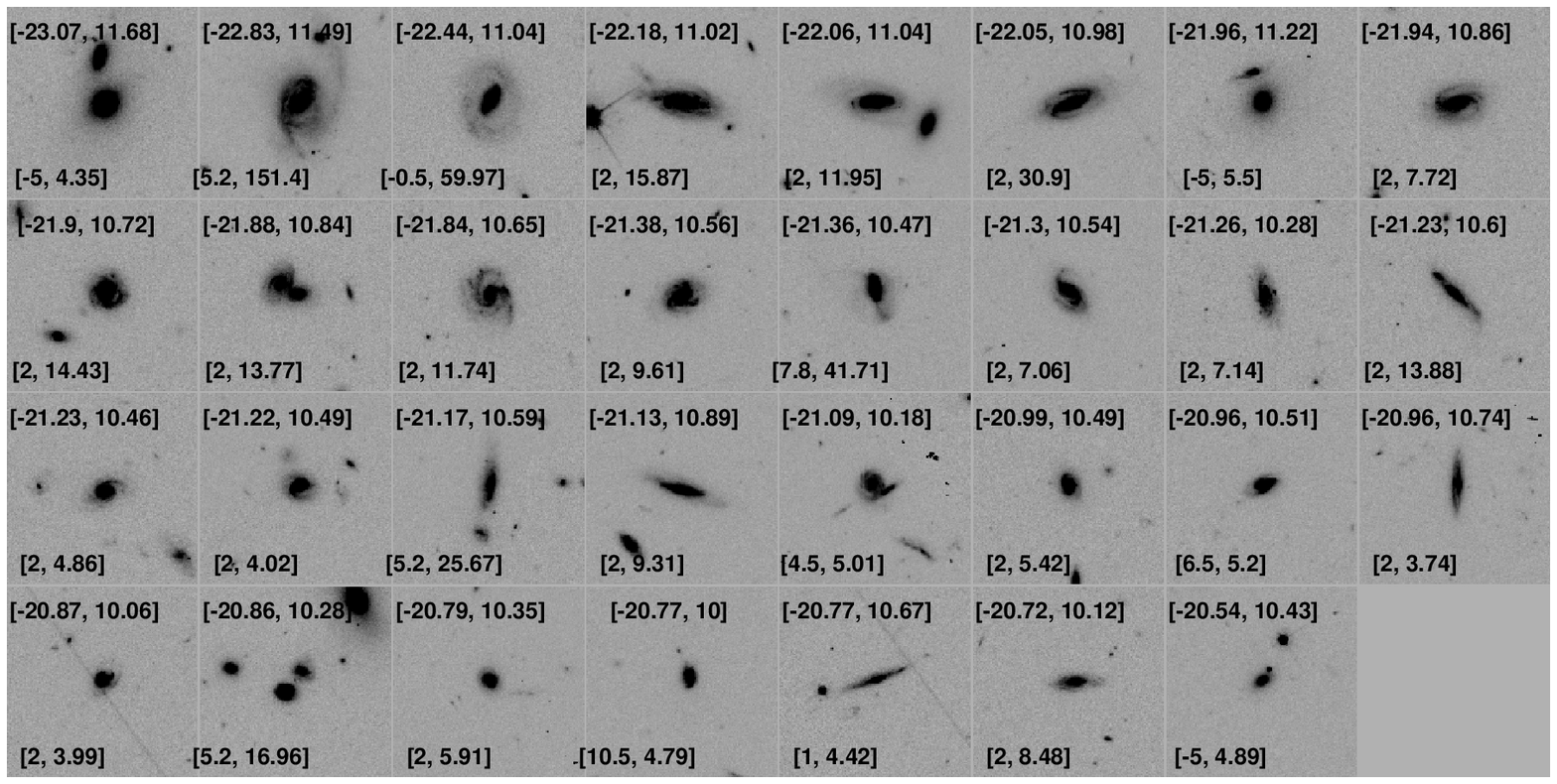}
\caption{Images ($15''\times15''$) of the supergroup galaxies
that fall on the HST/ACS mosaic ($M_V<$\magcut) with \sfrir.  Each
image lists the absolute $V$ magnitude (Vega) and estimated
$\log(M_{\ast})[M_{\odot}]$ on top, as well as the galaxy's Hubble
Type and \mipsmu~star formation rate on the bottom. Close
galaxy pairs (separation$<1''$) are considered single objects in both
the master $R$ catalog and MIPS catalog (see \S2.1 \&
\S\ref{mipstext}) because of the \mipsmu~imaging's resolution
(resampled pixel scale of 1.2$''$).
\label{tnails}}
\end{figure}

In the luminosity-limited sample, the fraction of \mipsmu~sources in
the cluster is significantly lower than in the field
(7$\pm2$\%\footnote[15]{Given the small number statistics, we assume a
binomial distribution to calculate the error on the fractions.}
vs. 36$\pm9$\%; Table~\ref{tab:fractions}).  However, we find that
{\it the fraction of \mipsmu~sources in the supergroup ($32\pm5$\%) is
comparable to the field and four times greater than in the cluster}.
Figure~\ref{tnails} shows HST/ACS images of the supergroup galaxies
with \sfrir~and $M_V<$\magcut.

Because ongoing star formation can increase a galaxy's total optical
luminosity and thus scatter lower-mass systems into the
luminosity-selected sample, we compare our results to a mass-selected
sample (Table~\ref{tab:fractions}).  As in Paper I, we consider only
galaxies with stellar masses greater than
$M_{\ast}=4\times10^{10}~M_{\odot}$.  Again, the fraction of
\mipsmu~sources in the supergroup is higher than in the cluster
(19$\pm5$\% vs. 5$\pm2$\%; Table~\ref{tab:fractions}); however, the
fraction in the supergroup is now only about half that of the field.

In applying a mass-cut, we discover that SG1120 has a significant
number of members (17; see Fig.~\ref{tnails}) that are bright
($M_V<$\magcut), mostly late-type galaxies with stellar masses of
($10.0< \log(M_{\ast})[M_{\odot}]<10.6$).  Once star formation is
quenched in these systems, they will fade and redden, and most will
have $L<L^{\ast}$, $i.e.$ they will populate the faint end of the red
sequence

We have assumed that the \mipsmu~emission is due to star formation but
as many authors have noted \citep[$e.g.$][]{donley:08}, there is a
likely contribution from dust-enshrouded active galactic nuclei (AGN).
However, we stress that AGN contamination does not impact our
conclusions because the relative fraction in each environment is
small.  In a 70 ksec $Chandra$/ACIS image of the SG1120 field
\citep{gonzalez:05}, only 4 of the 143 group galaxies are detected as
X-ray point sources.  As for the cluster galaxies ($z=0.33$),
\citet{martini:07} estimate that the AGN fraction in two $z\sim0.3$
clusters is less than 3\%.  The possible number of field AGN is
equally low: using \citet{donley:08}'s survey of IR-detected AGN, we
estimate that only one of the field IR sources can be an AGN.  Note
that if we account for these estimates of the AGN fraction, the
difference in the IR star-forming fraction between the supergroup and
cluster environment only increases ($\sim28$\% vs. $\sim4$\% in the
luminosity-selected samples).

\subsection{Infrared Luminosity Function}

\begin{figure}
\epsscale{1.0}
\plotone{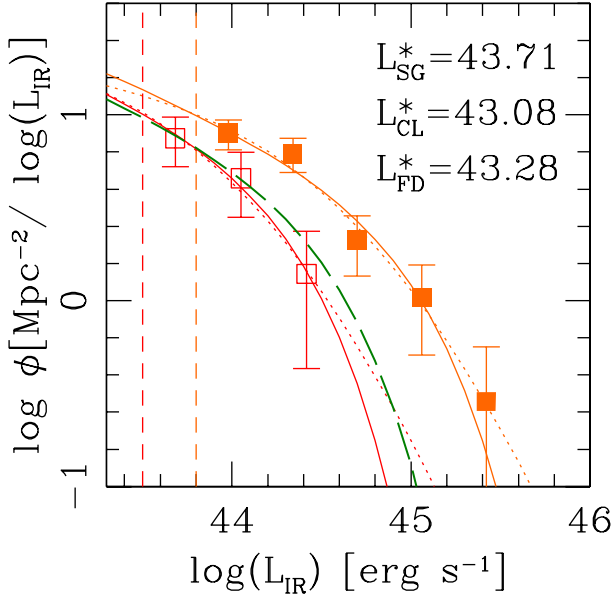}
\caption{The infrared luminosity functions (IR LFs) of the galaxy
groups ($z\sim0.37$; solid squares) and the galaxy cluster
($z=0.33$; open squares); $1\sigma$ errorbars are included.  The vertical
dashed lines at $\log(L_{IR})$[erg~s$^{-1}$]$=43.8$ and 43.5
correspond to the 80\% 
completeness limit of the \mipsmu~sources for the galaxy groups and
the cluster, respectively.  The IR LFs in both environments are well
fit by both Schechter (solid curves) and double-exponential (dotted
curves) functions.  The IR LF of the cluster galaxies is
consistent with the IR LF of $z\sim0$ clusters evolved to $z\sim0.35$
(long-dashed curve; see text for details), but the group IR LF has an
excess of sources.
\label{lumfunc}}
\end{figure}

To better quantify how the \mipsmu~sources in the supergroup differ
from their counterparts in the cluster and in the field, we compare
the infrared luminosity functions (IR LFs) of the supergroup (SG1120)
and cluster (CL1358) with published results from the field in
Fig.~\ref{lumfunc}.  We correct the observations in both SG1120 and
CL1358 for spectroscopic and \mipsmu~incompleteness; the 80\%
completeness limit for the \mipsmu~sources is
$\log(L_{IR})$[erg~s$^{-1}$]$=43.8$ and 43.5 in the supergroup and
galaxy cluster, respectively.

To fit the IR distribution, we follow \citet{bai:06} and first use a
Schechter function \citep{schechter:76}:

\begin{equation}
\phi(L) = \phi^{\ast}\left( \frac{L}{L^{\ast}} \right)^{-\alpha} \exp^{(-L/L^{\ast})}
\end{equation}

\noindent where we fix the faint-end slope $\alpha$ and adopt the
best-fit chi-square minimization method to determine $L^{\ast}$ and
$\phi^{\ast}$ (Table~\ref{tab:IRLF}).  Because studies show the IR LF
in general has a relatively large number of bright sources and is
better described by a double-exponential function \citep{lefloch:05},
we adopt their approach and also fit a double-exponential function:

\begin{equation}
\phi(L) = \phi^{\ast} \left( \frac{L}{L^{\ast}} \right)^{(1-\alpha)}
\exp \left\{ -\left( \frac{1}{2\sigma^2} \right) 
\log ^2 \left[1+\left(\frac{L}{L^{\ast}}\right) \right] \right\}
\end{equation}

\noindent where we fix the constants $\alpha$ and $\sigma$ to the
values measured for the field IR LF, and minimize with chi-square
again.  Note that to determine the faint-end slope $\alpha$ in either
function, deeper IR observations are required \citep{bai:06}.

The IR LFs in both the supergroup and galaxy cluster are well-fit by
both a Schechter and a double-exponential function
(Fig.~\ref{lumfunc}) using different values for $L^{\ast}$ and
$\phi^{\ast}$ (see Table~\ref{tab:IRLF}).  However, the density of IR
sources in the supergroup is dramatically higher than in the cluster,
especially at $\log(L_{IR})$[erg~s$^{-1}$]$>45$.

Perhaps the large difference is due to CL1358 being unusually
deficient in IR sources.  We test this by taking the IR LF determined
from \mipsmu~observations of galaxy clusters at $z\sim0$
\citep{bai:06} and evolve the IR LF to $z\sim0.35$ using the observed
evolution in the field IR LF \citep{lefloch:05}.
\citet{bai:07,bai:09} find that $L_{IR}^{\ast}$ and
$\phi_{IR}^{\ast}$, as derived from \mipsmu~observations, evolve in
approximately the same manner in galaxy clusters and the field to
$z\sim0.8$, although see \citet{muzzin:08} for an alternative result.
The IR LF for CL1358 ($z=0.33$) is consistent with the evolved cluster
IR LF (Fig.~\ref{lumfunc}; long-dashed curve).  However, the density
of IR sources in the supergroup is $\sim10$ times higher than the
number predicted from the evolved IR LF at
$\log(L_{IR})$[erg~s$^{-1}$]$>45$.

The IR LF in the supergroup also differs from the IR LF for the field
measured by \citet{lefloch:05}.  The best-fit double-exponential
function to the group IR LF has a measurably larger $L_{IR}^{\ast}$
compared to the value for field galaxies at ($0.3<z<0.45$):
$43.71^{+0.19}_{-0.19}$ vs. $43.28^{+0.09}_{-0.03}$.  In comparison,
\citet{bai:09} find a similar $L_{IR}^{\ast}$ value for both local
cluster and field galaxies.  (Table~\ref{tab:IRLF}).  SG1120's higher
$L_{IR}^{\ast}$ relative to even that measured for the field suggests
that star formation is enhanced in the group environment.

To summarize, the number of IR sources in the galaxy groups that make
up SG1120 is significantly higher than in CL1358, a rich galaxy
cluster at $z=0.33$, and includes a population of very bright IR
sources ($\log(L_{IR})$[erg~s$^{-1}$]$>45$) that are not found in
CL1358 nor in lower redshift clusters.  The higher value of
$\log(L^{\ast}_{IR})$ in the supergroup compared to that measured for
the field at $(0.3<z<0.45)$ also indicates that the IR sources in the
supergroup differ from their counterparts in the field, $i.e.$ that
star formation likely is enhanced in the group environment.

\subsection{Local Environment}\label{environment}

Having established that the \mipsmu~population in the supergroup
(SG1120) is different from that in the cluster (CL1358) and likely
also the field environment, we examine how the galaxy populations for
the luminosity-selected samples ($M_V<$\magcut) depend on local
environment, $i.e.$ how star formation rate relates to galaxy density
\citep{balogh:98,gomez:03}.  In addition to the IR-bright population,
we separate galaxies into optically-defined absorption-line
([OII]$\lambda3727<5$\AA) and emission-line
([OII]$\lambda3727\geq5$\AA) systems. In the supergroup and cluster,
we define the local galaxy density $\Sigma$ using the distance to the
$10^{th}$ nearest spectroscopically confirmed neighbor; we note that
the following results do not change if we use instead the $5^{th}$
nearest neighbor.

\begin{figure} 
\epsscale{1.0} 
\plotone{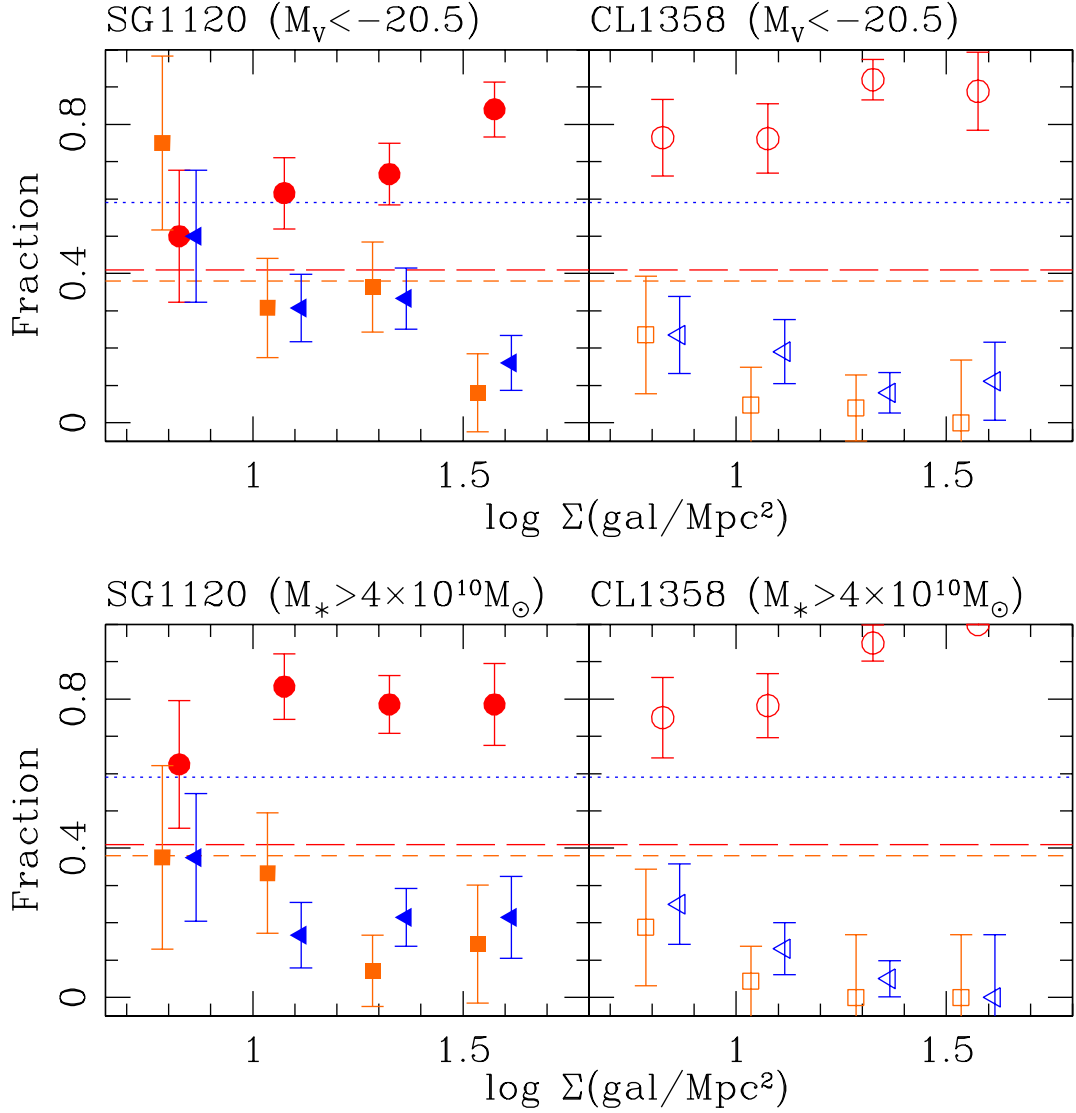} 
\caption{Relative fraction of absorption-line (circles), emission-line
(triangles), and \mipsmu~(squares) members as a function of local
galaxy surface density in the groups (SG1120, left panels, filled
symbols) and cluster (CL1358, right panels, open symbols); the points
are offset slightly in $\log\Sigma$ for clarity.  The top panels show
the luminosity-selected ($M_V<$\magcut) members and the bottom panels
the mass-selected ($M_{\ast}>$\masscut) members.  The long-dashed,
dotted, and short-dashed horizontal lines show respectively the
absorption-line, emission-line, and \mipsmu~fractions in the
field. Only in the supergroup with a luminosity-selected sample does
the \mipsmu~fraction increase steadily with decreasing local density,
$i.e.$ an infrared star formation--density relation.
\label{fdensity}}
\end{figure}

Figure~\ref{fdensity} (top left) shows how the fraction of
\sfrir~members in the supergroup steadily increases with decreasing
local density.  The increasing fraction of emission-line members with
decreasing $\Sigma$ mirrors the trend for IR members, and the
absorption-line fraction changes accordingly.  The trend of an
increasing IR fraction with decreasing local density remains even if
we apply higher IR star formation rate threshold of 5
$M_{\odot}$~yr$^{-1}$.  In contrast, the IR population in the cluster
(top right) shows essentially no trend with local environment: the
absorption-line population dominates throughout the range of local
densities explored here ($5<\Sigma<70$ gal~Mpc$^{-2}$)\footnote[16]{While
the galaxy density in the cluster environment extends to
$\Sigma>100$~gal~Mpc$^{-2}$, we consider only the $\Sigma$ range that
overlaps with the galaxy groups.}.  These results are in line with
Paper I where we also find an increase in \mipsmu~members outside the
cores of massive clusters ($R_{proj}>700$\hi~kpc).  Our results argue
for a physical mechanism that quenches star formation before the
members reach the group cores.

At the lowest galaxy densities, the fraction of IR members in the
supergroup is higher than even in the field: considering all members
with $\Sigma<20$ gal~Mpc$^{-2}$, the IR fraction increases to 49\%
(26/53) compared to the field value of 38\%
(Table~\ref{tab:fractions}).  At these low galaxy densities, the
higher fraction of IR members in the supergroup relative to the field,
while statistically not significant, is consistent with the higher
$L^{\ast}_{IR}$ measured in the group environment (see
Table~\ref{tab:IRLF}).

If we now examine how the mass-selected ($M_{\ast}>$\masscut) samples
depend on local environment (Fig.~\ref{fdensity}, bottom panels), the
trend of increasing \mipsmu~fraction with decreasing galaxy density in
the supergroup is weaker: the absorption-line population dominates in
both the group and the cluster environment at
$\Sigma>10$~gal~Mpc$^{-2}$.  These results are consistent with H07 who
find that evolution in the early-type fraction in massive clusters
($0<z<0.8$) is weaker when considering only galaxies with
$M_{\ast}>$\masscut~versus a luminosity-selected sample, thus galaxies
with lower masses play a significant role in the observed evolution of
the cluster galaxy population.

We find that the supergroup has a population of luminous,
\mipsmu~detected late-type members with stellar masses of
($10.0<\log(M_{\ast})[M_{\odot}]<10.6$) that are mostly outside the
groups' cores (see Figures~\ref{sg1120-xy} \& \ref{tnails}), $i.e.$ at
lower galaxy densities.  It is these galaxies that cause the strong
observed trend of decreasing \mipsmu~fraction with increasing galaxy
density in the {\it luminosity-selected} sample.  As noted in
\S\ref{mipsfrac}, most of these will fade and redden and can populate
the faint end of the red sequence once their star formation is
quenched.

%%%%%%%%%%%%%%%%%%%%%%%%%%%%%%%%%%%%%%%%%%%%%%%%%%%%%%%%%%%%%%%%%%%%
\section{Discussion}

% colors, stellar masses, morphologies

To study how galaxies evolve, galaxies are usually separated into
active/emission-line and passive/absorption-line systems with the goal
of isolating the physical process that connects the two phases, $e.g.$
removal of a galaxy's gas halts its star formation and the galaxy
evolves from an active system into a passive one
\citep{abadi:99,kawata:08}.  The high fraction of \mipsmu~galaxies in
the supergroup and the field means that obscured star formation (IR phase)
is important for at least 30\% of optically-selected galaxies in both
these environments; the IR phase is likely to be as important in
clusters given that clusters grow via the accretion of field and group
galaxies \citep{peebles:70}.  In the following, we examine the
physical properties of the \mipsmu~galaxies to better understand where
the IR population fits into our current picture of galaxy evolution.

\subsection{Morphologies of \mipsmu~Galaxies}

To determine what the typical morphology of a \mipsmu~galaxy is across
environment at $z\sim0.35$, we separate the samples into late-type
($T>0$; disk-dominated) and early-type ($T\leq0$; bulge-dominated)
systems.  In both the field and supergroup, most ($>60$\%) of the
late-type galaxies are IR-detected; the supergroup even has a few
bulge-dominated systems that are IR-detected\footnote[17]{While none of
the eight early-types in the field have \sfrir, this is likely due to
our relatively small field sample because \citet{lotz:08} do find a
number of early-type galaxies at $(0.4<z<1.2)$ with comparable IR
luminosities.}.  In contrast, only $\sim30$\% of the late-type
galaxies in the cluster are IR-detected, and none of the cluster's
bulge-dominated members are IR-detected.  These differences are true
in both the luminosity and mass-selected samples
(Table~\ref{tab:fractions}).

Comparing the \mipsmu~galaxies in the supergroup to the field again
strongly suggests a difference between the two environments.  In the
luminosity-selected samples, the supergroup and field have similar
\mipsmu~fractions.  However, the supergroup has a much higher fraction
of E/S0 members: the E/S0 fraction in the supergroup is $\sim60$\% but
it is only $\sim30$\% in the field\footnote[18]{The E/S0 fraction in our
field sample is consistent with the E/S0 fraction measured by
\citet{driver:98} for a significantly larger field sample.}
(Table~\ref{tab:fractions}).  Several of the \mipsmu~galaxies in the
supergroup are in merging, disk-dominated systems (see
Fig.~\ref{tnails}), but only one of the massive dissipationless
merging pairs \citep{tran:08} is detected at \mipsmu.

Despite having double the fraction of early-type galaxies compared to
the field, the supergroup has a high \mipsmu~fraction due to a
population of luminous ($M_V<$\magcut), low-mass ($M_{\ast}<$\masscut)
late-type members with \sfrir~(see \S\ref{mipsfrac},
\S\ref{environment}, \& Table~\ref{tab:fractions}).  Note that in the
mass-selected sample, the \mipsmu~fraction in the supergroup drops
from $\sim32$\% to $\sim19$\% while the field fraction remains high
($\sim35$\%).  Our results show that the timescales for morphological
evolution and quenching of star formation must differ (see also Finn
et al., in prep).

\subsection{Star Formation on the Red Sequence}

Across all three environments, there are \mipsmu~sources that are also
on the optical red sequence (see Fig.~\ref{cmd}); here we use the
classical definition of the red sequence as galaxies with
$\Delta(B-V)>-0.2$ \citep{butcher:84}.  The fraction of red
\mipsmu~sources depends on environment: it is highest in the field
(21\%; 6/29), decreases in the supergroup (7\%; 7/98), and is is
lowest in the cluster (3\%; 3/105).  These results do not depend on
whether we use the luminosity or mass-selected sample.

Our results appear to conflict with \citet{gallazzi:08} who find that
the fraction of red \mipsmu~sources in the Abell 901/902 supercluster
\citep[$z=0.165$;][]{gray:02} peaks at intermediate densities typical
of cluster outskirts and galaxy groups\footnote[19]{Because
\citet{gallazzi:08} estimate local galaxy density differently, we
cannot compare their values directly to Fig.~\ref{fdensity}.  These
authors use a spectroscopic sample supplemented with members selected
with photometric redshifts.}.  However, the authors estimate the field
contamination in their magnitude range can be as high as 20\%, and our
study shows that the fraction of red \mipsmu~galaxies in the field is
$\sim3$ times higher than in the groups.  By using a spectroscopically
selected sample, we circumvent possible problems due to field
contamination.

In A901/902, \citet{wolf:09} find that the dusty red star-forming
members are primarily spiral galaxies, and that this population mostly
overlaps with the ``optically passive'' spirals (as defined by color).
We find similar results: all of the red \mipsmu~galaxies in the field
and cluster are disk-dominated systems ($T>0$), and most of the red
\mipsmu~galaxies ($\sim60$\%) in the groups are spirals as well (see
Fig.~\ref{tnails}). 

We note that neither optical colors nor optical spectroscopy reliably
identifies dusty red [$\Delta(B-V)>-0.2$] star-forming spirals:
summing across environment, only (10/15) of the red \mipsmu~galaxies
have [OII]$\lambda3727>5$\AA, $i.e.$ one third of \mipsmu~members on
the red sequence show no significant [OII] emission.  This result is
in line with earlier studies, $e.g.$ \citet{moustakas:06}, that show
optical spectroscopy can severely underestimate the level of activity.
On a related note, many \mipsmu~galaxies can be strongly extincted
with E(B-V) values as high as $0.6$ \citep{cowie:08}; once corrected
for extinction, many of the \mipsmu~galaxies would not lie on the red
sequence.

\subsection{Progenitors of Faint Red Galaxies}

\begin{figure} 
\epsscale{1.0} 
\plotone{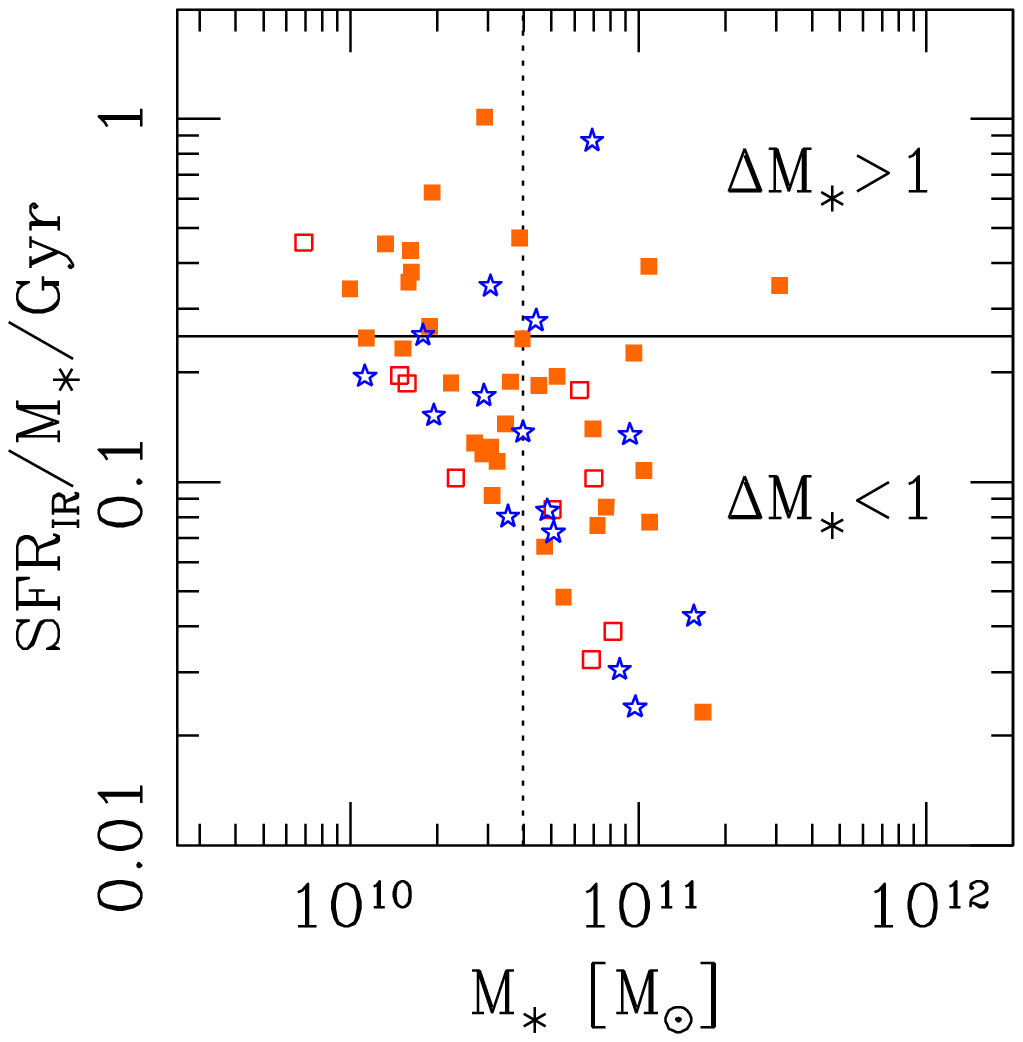} 
\caption{The \mipsmu-derived specific star formation rate (SSFR) in units
of Gyr for the galaxy groups (filled squares), cluster (open squares), and
the field sample (stars); the dotted vertical line denotes
$M_{\ast}=$\masscut~and we show only members with \sfrir.  If we assume
constant SF rates, the galaxies above the horizontal line will more
than double their stellar mass from $z\sim0.37$ to now.  At most,
perhaps four of the 72 massive galaxies in SG1120 can double their
stellar mass.  However, there are a number of lower-mass
($10.0<\log(M_{\ast})[M_{\odot}]<10.6$) group galaxies that can grow
substantially in stellar mass (upper left quadrant); these are
luminous late-type members with \sfrir. 
\label{ssfr}} 
\end{figure}

As the groups in SG1120 merge to form a galaxy cluster, how do the
\mipsmu~members impact the overall galaxy population?  In
Fig.~\ref{ssfr}, we plot specific star formation rates (defined as
\sfrir~divided by stellar mass) versus stellar mass for the
\mipsmu~galaxies in the supergroup, field, and cluster.  Assuming the
\mipsmu~members maintain their current star formation rates, perhaps
only five out of the 72 massive ($M_{\ast}>$\masscut) group galaxies
will double their stellar masses, $i.e.$ virtually all of the massive
galaxies that will end up in the cluster are already in place.

In our analysis, we have identified a considerable number of luminous
($M_V<$\magcut) galaxies in the supergroup that have \sfrir~and
stellar masses below \masscut~(17/98; Table~\ref{tab:fractions}); it
is this population that contributes the most to the difference between
the \mipsmu~population in the supergroup and in the cluster.
Fig.~\ref{ssfr} shows that even if these group galaxies can maintain
their current star formation rates, most (15/17; $\sim90$\%) will
still have stellar masses of $M_{\ast}<10^{11}M_{\odot}$~at $z\sim0$;
the current average stellar mass for all 17 galaxies is
$\log(M_{\ast})[M_{\odot}]=10.4$.  Note that most of these members are
at $R_{proj}>0.5$~Mpc from their respective group cores (see
Fig.~\ref{sg1120-xy}). 

We test our hypothesis that these galaxies can evolve into
($L<L^{\ast}$) red galaxies by comparing their stellar masses to the
faint red galaxies in CL1358, our massive galaxy cluster.  Following
\citet{delucia:07a}, we define faint red galaxies as having
luminosities of $(0.1-0.4)L^{\ast}$\footnote[20]{Using $M_V^{\ast}=-20.6$,
the corresponding Vega $V$-band magnitudes are $(-20.1<M_V<19.6)$.}
and $\Delta(B-V)>-0.2$.  The average stellar mass of the faint red
galaxies in the cluster is $\log(M_{\ast})[M_{\odot}]=10.3$; this is
comparable to the average stellar mass of the luminous, low-mass,
\sfrir~supergroup galaxies.  Assuming their star formation is quenched
by $z\sim0$, these supergroup galaxies will fade and redden to lie on
the CM relation in less than a Gyr \citep[see models
by][]{bruzual:03}.  Their younger luminosity-weighted ages relative to
the more massive galaxies will be consistent with the observed age
spread in the Coma cluster \citep{poggianti:01b}.  We stress that the
luminous, low-mass, \sfrir~supergroup galaxies are likely to be only
one of multiple progenitors of faint red galaxies.

%%%%%%%%%%%%%%%%%%%%%%%%%%%%%%%%%%%%%%%%%%%%%%%%%%%%%%%%%%%%%%%%%%%%
\section{Summary}

To quantify how dust-obscurred star formation varies with environment,
we compare galaxies in a super galaxy group to those in the field and in
a massive cluster at $z\sim0.35$ using a rich multi-wavelength dataset
that includes imaging from $Hubble$ (optical), $Chandra$ (X-ray), and
$Spitzer$ (\mipsmu).  The strength of our work relies on extensive
optical spectroscopy in our fields: the magnitude-limited
spectroscopic surveys yielded a total of over 1800 unique redshifts
and enable us to securely identify field, supergroup, and cluster
members. We focus on the four X-ray luminous galaxy groups at $z\sim0.37$
(SG1120-12) that will merge to form a galaxy cluster comparable in
mass to Coma \citep{gonzalez:05}; the groups have line-of-sight
velocity dispersions of $303-580$\kms.  To ensure robust comparison,
we consider only field galaxies at $0.25\leq z\leq0.45$
($\bar{z}=0.35$) and confirmed members of the massive galaxy cluster
CL1358+62 \citep[$z=0.33$,][]{fisher:98}.

We find that the supergroup has a significantly higher fraction of
dusty star-forming members than the massive galaxy cluster: in the
luminosity-selected ($M_V<$\magcut) samples, 32\% of the supergroup
members have \sfrir~compared to only 7\% of the cluster members.  The
supergroup's infrared luminosity function confirms that the density of
IR sources is dramatically higher in the groups compared to the
cluster.  The supergroup members also include bright IR sources
($\log(L_{IR})$[erg~s$^{-1}$]$>45$) not found in galaxy clusters at
$z\lesssim0.35$.

When selected by luminosity, the supergroup members show a strong
trend of decreasing \mipsmu~fraction with increasing local galaxy
density, $i.e.$ an infrared-density relationship.  This mirrors the
trend in the optically active members (as defined by [OII] emission).
In contrast, the fraction of \mipsmu~sources in the massive cluster
stays essentially zero at all densities.

Comparison to the mass-selected ($M_{\ast}>$\masscut) samples reveals
that the higher \mipsmu~fraction and the IR-density relation in the
supergroup is due primarily to a population of luminous
($M_V<$\magcut), lower-mass ($10.0<\log(M_{\ast})[M_{\odot}]<10.6$),
late-type members with \sfrir~($\sim17$\%).  Most of these members are
outside of the group cores ($R_{proj}\geq0.5$~Mpc).  Assuming their
star formation is quenched in the next $\sim3-4$ Gyr, these members
will fade and redden by $z\sim0$, and most will become fainter
($L<L^{\ast}$) galaxies on the color-magnitude relation. The physical
mechanism that quenches their star formation must be effective outside
the group cores, $i.e.$ in lower density environments.

In the supergroup, the excess of \mipsmu~sources, the number of very
bright \mipsmu~members, and the infrared-density relationship is
surprising because the E/S0 fraction is already as high as in the
cluster \citep[$>60$\% for luminosity-selected sample;][]{kautsch:08}.
No further morphological evolution is required to bring the
morphological distribution of the groups in line with the high
early-type fractions observed in local galaxy clusters.  In other
words, the timescale for morphological transformation must not be
strongly coupled to when star formation is completely quenched.  

Our analysis indicates that the \mipsmu~population in the supergroup
differs even from the field: 1) the supergroup's IR luminosity
function has a measurably higher $L_{IR}^{\ast}$ than the field; and
2) the E/S0 fraction in the supergroup is twice that of the field, yet
the \mipsmu~fraction in both environments are comparable.  If dusty
star formation is enhanced in the supergroup relative to the field,
our IR-density analysis suggests that it occurs at densities of
$\Sigma<20$~gal~Mpc$^{-2}$.  A larger field sample selected with the
same criteria as in the supergroup and cluster is needed to answer this
question securely.

Our study highlights the importance of understanding galaxy evolution
on group scales.  A significant fraction ($\gtrsim30$\%) of optically
selected galaxies in both the supergroup and field at $z\sim0.35$ have
dust-obscured star formation; the IR-phase must be as important in
clusters because clusters grow by accreting galaxy groups and field
galaxies. As demonstrated in recent simulations of galaxy groups
\citep[$e.g.$][]{hester:06,romeo:08,mccarthy:08,kawata:08}, the
physical mechanisms that affect star formation and induce
morphological evolution are already well underway in the galaxy groups
that make up SG1120.  We will continue dissecting how these galaxies
are transformed by using recently obtained IFU observations to map the
kinematics and star formation of the \mipsmu~members.

%%%%%%%%%%%%%%%%%%%%%%%%%%%%%%%%%%%%%%%%%%%%%%%%%%%%%%%%%%%%%%%%%%%%

\acknowledgments

K.T. and A.S. acknowledge support from the Swiss National Science
Foundation (grant PP002-110576).  J.M. acknowledges support from
NASA-06-GALEX06-0030 and Spitzer G05-AR-50443, and L.B. from NASA
Spitzer programs through JPL subcontracts \#1255094 and \#1256318.
Support was also provided by NASA HST G0-10499, JPL/Caltech SST
GO-20683, and Chandra GO2-3183X3.  

{\it Facilities:} VLT (VIMOS), VLT (FORS2), Magellen (LDSS3), KPNO
(Mayall 4m), HST (ACS), SST (MIPS), CXO (ACIS), Keck (LRIS).

\bibliographystyle{/Users/vy/aastex/apj}
\bibliography{/Users/vy/aastex/tran}

\begin{thebibliography}{101}
\expandafter\ifx\csname natexlab\endcsname\relax\def\natexlab#1{#1}\fi

\bibitem[{{Abadi} {et~al.}(1999){Abadi}, {Moore}, \& {Bower}}]{abadi:99}
{Abadi}, M.~G., {Moore}, B., \& {Bower}, R.~G. 1999, \mnras, 308, 947

\bibitem[{{Abazajian} {et~al.}(2003){Abazajian}, {Adelman-McCarthy},
  {Ag{\"u}eros}, {Allam}, {Anderson}, {Annis}, {Bahcall}, \& et~al.}]{sdss1:03}
{Abazajian}, K., {Adelman-McCarthy}, J.~K., {Ag{\"u}eros}, M.~A., {Allam},
  S.~S., {Anderson}, S.~F., {Annis}, J., {Bahcall}, N.~A., \& et~al. 2003, \aj,
  126, 2081

\bibitem[{{Abell}(1977)}]{abell:77}
{Abell}, G.~O. 1977, \apj, 213, 327

\bibitem[{{Appenzeller} {et~al.}(1998){Appenzeller}, {Fricke}, {F{\"u}rtig},
  {G{\"a}ssler}, {H{\"a}fner}, {Harke}, {Hess}, \& et~al.}]{appenzeller:98}
{Appenzeller}, I., {Fricke}, K., {F{\"u}rtig}, W., {G{\"a}ssler}, W.,
  {H{\"a}fner}, R., {Harke}, R., {Hess}, H.-J., \& et~al. 1998, The Messenger,
  94, 1

\bibitem[{{Bai} {et~al.}(2007){Bai}, {Marcillac}, {Rieke}, {Rieke}, {Tran},
  {Hinz}, {Rudnick}, {Kelly}, \& {Blaylock}}]{bai:07}
{Bai}, L., {Marcillac}, D., {Rieke}, G.~H., {Rieke}, M.~J., {Tran}, K.-V.~H.,
  {Hinz}, J.~L., {Rudnick}, G., {Kelly}, D.~M., \& {Blaylock}, M. 2007, \apj,
  664, 181

\bibitem[{{Bai} {et~al.}(2008){Bai}, {Rieke}, {Rieke}, {Christlein}, \&
  {Zabludoff}}]{bai:09}
{Bai}, L., {Rieke}, G.~H., {Rieke}, M.~J., {Christlein}, D., \& {Zabludoff},
  A.~I. 2008, ArXiv e-prints

\bibitem[{{Bai} {et~al.}(2006){Bai}, {Rieke}, {Rieke}, {Hinz}, {Kelly}, \&
  {Blaylock}}]{bai:06}
{Bai}, L., {Rieke}, G.~H., {Rieke}, M.~J., {Hinz}, J.~L., {Kelly}, D.~M., \&
  {Blaylock}, M. 2006, \apj, 639, 827

\bibitem[{{Balogh} {et~al.}(2009){Balogh}, {McGee}, {Wilman}, {Bower}, \&
  et~al}]{balogh:09}
{Balogh}, M.~L., {McGee}, S.~L., {Wilman}, D., {Bower}, R.~G., \& et~al. 2009,
  ArXiv e-prints

\bibitem[{{Balogh} {et~al.}(1998){Balogh}, {Schade}, {Morris}, {Yee},
  {Carlberg}, \& {Ellingson}}]{balogh:98}
{Balogh}, M.~L., {Schade}, D., {Morris}, S.~L., {Yee}, H.~K.~C., {Carlberg},
  R.~G., \& {Ellingson}, E. 1998, \apjl, 504, L75

\bibitem[{{Baum} {et~al.}(1997){Baum}, {Hammergren}, {Thomsen}, {Groth},
  {Faber}, {Grillmair}, \& {Ajhar}}]{baum:97}
{Baum}, W.~A., {Hammergren}, M., {Thomsen}, B., {Groth}, E.~J., {Faber}, S.~M.,
  {Grillmair}, C.~J., \& {Ajhar}, E.~A. 1997, \aj, 113, 1483

\bibitem[{Beers {et~al.}(1990)Beers, Flynn, \& Gebhardt}]{beers:90}
Beers, T.~C., Flynn, K., \& Gebhardt, K. 1990, \aj, 100, 32

\bibitem[{{Bekki} {et~al.}(2002){Bekki}, {Couch}, \& {Shioya}}]{bekki:02}
{Bekki}, K., {Couch}, W.~J., \& {Shioya}, Y. 2002, \apj, 577, 651

\bibitem[{{Bell} \& {de Jong}(2001)}]{bell:01}
{Bell}, E.~F. \& {de Jong}, R.~S. 2001, \apj, 550, 212

\bibitem[{{Bell} {et~al.}(2003){Bell}, {McIntosh}, {Katz}, \&
  {Weinberg}}]{bell:03b}
{Bell}, E.~F., {McIntosh}, D.~H., {Katz}, N., \& {Weinberg}, M.~D. 2003, \apjs,
  149, 289

\bibitem[{{Bertin}(2006)}]{bertin:06}
{Bertin}, E. 2006, in Astronomical Society of the Pacific Conference Series,
  Vol. 351, Astronomical Data Analysis Software and Systems XV, ed.
  C.~{Gabriel}, C.~{Arviset}, D.~{Ponz}, \& S.~{Enrique}, 112--+

\bibitem[{{Bertin} \& {Arnouts}(1996)}]{bertin:96}
{Bertin}, E. \& {Arnouts}, S. 1996, \aaps, 117, 393

\bibitem[{{Bertin} {et~al.}(2002){Bertin}, {Mellier}, {Radovich}, {Missonnier},
  {Didelon}, \& {Morin}}]{bertin:02}
{Bertin}, E., {Mellier}, Y., {Radovich}, M., {Missonnier}, G., {Didelon}, P.,
  \& {Morin}, B. 2002, in Astronomical Society of the Pacific Conference
  Series, Vol. 281, Astronomical Data Analysis Software and Systems XI, ed.
  D.~A. {Bohlender}, D.~{Durand}, \& T.~H. {Handley}, 228--+

\bibitem[{{Blanton} \& {Berlind}(2007)}]{blanton:07b}
{Blanton}, M.~R. \& {Berlind}, A.~A. 2007, \apj, 664, 791

\bibitem[{{Blanton} \& {Roweis}(2007)}]{blanton:07a}
{Blanton}, M.~R. \& {Roweis}, S. 2007, \aj, 133, 734

\bibitem[{{Bower} {et~al.}(1992){Bower}, {Lucey}, \& {Ellis}}]{bower:92}
{Bower}, R.~G., {Lucey}, J.~R., \& {Ellis}, R.~S. 1992, \mnras, 254, 601+

\bibitem[{{Bruzual} \& {Charlot}(2003)}]{bruzual:03}
{Bruzual}, G. \& {Charlot}, S. 2003, \mnras, 344, 1000

\bibitem[{{Butcher} \& {Oemler}(1978)}]{butcher:78}
{Butcher}, H. \& {Oemler}, A. 1978, \apj, 219, 18

\bibitem[{{Butcher} \& {Oemler}(1984)}]{butcher:84}
---. 1984, \apj, 285, 426

\bibitem[{{Cooper} {et~al.}(2008){Cooper}, {Newman}, {Weiner}, {Yan},
  {Willmer}, {Bundy}, \& et~al.}]{cooper:08}
{Cooper}, M.~C., {Newman}, J.~A., {Weiner}, B.~J., {Yan}, R., {Willmer},
  C.~N.~A., {Bundy}, K., \& et~al. 2008, \mnras, 383, 1058

\bibitem[{{Cowie} \& {Barger}(2008)}]{cowie:08}
{Cowie}, L.~L. \& {Barger}, A.~J. 2008, \apj, 686, 72

\bibitem[{{Crawford} {et~al.}(2009){Crawford}, {Bershady}, \&
  {Hoessel}}]{crawford:09}
{Crawford}, S.~M., {Bershady}, M.~A., \& {Hoessel}, J.~G. 2009, \apj, 690, 1158

\bibitem[{{Dale} {et~al.}(2007){Dale}, {Gil de Paz}, {Gordon}, {Hanson},
  {Armus}, {Bendo}, {Bianchi}, {Block}, \& et~al.}]{dale:07}
{Dale}, D.~A., {Gil de Paz}, A., {Gordon}, K.~D., {Hanson}, H.~M., {Armus}, L.,
  {Bendo}, G.~J., {Bianchi}, L., {Block}, M., \& et~al. 2007, \apj, 655, 863

\bibitem[{{Dale} \& {Helou}(2002)}]{dale:02}
{Dale}, D.~A. \& {Helou}, G. 2002, \apj, 576, 159

\bibitem[{{De Lucia} {et~al.}(2007){De Lucia}, {Poggianti},
  {Arag{\'o}n-Salamanca}, \& {et al.}}]{delucia:07a}
{De Lucia}, G., {Poggianti}, B.~M., {Arag{\'o}n-Salamanca}, A., \& {et al.}
  2007, \mnras

\bibitem[{{Donley} {et~al.}(2008){Donley}, {Rieke}, {P{\'e}rez-Gonz{\'a}lez},
  \& {Barro}}]{donley:08}
{Donley}, J.~L., {Rieke}, G.~H., {P{\'e}rez-Gonz{\'a}lez}, P.~G., \& {Barro},
  G. 2008, \apj, 687, 111

\bibitem[{Dressler(1980)}]{dressler:80}
Dressler, A. 1980, \apj, 236, 351

\bibitem[{{Dressler} {et~al.}(2009){Dressler}, {Rigby}, {Oemler}, {Fritz},
  {Poggianti}, {Rieke}, \& {Bai}}]{dressler:09}
{Dressler}, A., {Rigby}, J., {Oemler}, A., {Fritz}, J., {Poggianti}, B.~M.,
  {Rieke}, G., \& {Bai}, L. 2009, \apj, 693, 140

\bibitem[{{Driver} {et~al.}(1998){Driver}, {Fernandez-Soto}, {Couch},
  {Odewahn}, {Windhorst}, {Phillips}, {Lanzetta}, \& {Yahil}}]{driver:98}
{Driver}, S.~P., {Fernandez-Soto}, A., {Couch}, W.~J., {Odewahn}, S.~C.,
  {Windhorst}, R.~A., {Phillips}, S., {Lanzetta}, K., \& {Yahil}, A. 1998,
  \apjl, 496, L93+

\bibitem[{{Eastman} {et~al.}(2007){Eastman}, {Martini}, {Sivakoff}, {Kelson},
  {Mulchaey}, \& {Tran}}]{eastman:07}
{Eastman}, J., {Martini}, P., {Sivakoff}, G., {Kelson}, D.~D., {Mulchaey},
  J.~S., \& {Tran}, K.-V. 2007, \apjl, 664, L9

\bibitem[{{Elbaz} {et~al.}(2007){Elbaz}, {Daddi}, {Le Borgne}, {Dickinson},
  {Alexander}, {Chary}, {Starck}, \& et~al.}]{elbaz:07}
{Elbaz}, D., {Daddi}, E., {Le Borgne}, D., {Dickinson}, M., {Alexander}, D.~M.,
  {Chary}, R.-R., {Starck}, J.-L., \& et~al. 2007, \aap, 468, 33

\bibitem[{{Ellingson} {et~al.}(2001){Ellingson}, {Lin}, {Yee}, \&
  {Carlberg}}]{ellingson:01}
{Ellingson}, E., {Lin}, H., {Yee}, H.~K.~C., \& {Carlberg}, R.~G. 2001, \apj,
  547, 609

\bibitem[{Fabricant {et~al.}(2000)Fabricant, Franx, \& van
  Dokkum}]{fabricant:00}
Fabricant, D., Franx, M., \& van Dokkum, P. 2000, \apj, 539, 577

\bibitem[{{Fisher} {et~al.}(1998){Fisher}, {Fabricant}, {Franx}, \& {van
  Dokkum}}]{fisher:98}
{Fisher}, D., {Fabricant}, D., {Franx}, M., \& {van Dokkum}, P. 1998, \apj,
  498, 195+

\bibitem[{{G{\' o}mez} {et~al.}(2003){G{\' o}mez}, {Nichol}, {Miller},
  {Balogh}, {Goto}, {Zabludoff}, {Romer}, {Bernardi}, {Sheth}, {Hopkins},
  {Castander}, {Connolly}, {Schneider}, {Brinkmann}, {Lamb}, {SubbaRao}, \&
  {York}}]{gomez:03}
{G{\' o}mez}, P.~L., {Nichol}, R.~C., {Miller}, C.~J., {Balogh}, M.~L., {Goto},
  T., {Zabludoff}, A.~I., {Romer}, A.~K., {Bernardi}, M., {Sheth}, R.,
  {Hopkins}, A.~M., {Castander}, F.~J., {Connolly}, A.~J., {Schneider}, D.~P.,
  {Brinkmann}, J., {Lamb}, D.~Q., {SubbaRao}, M., \& {York}, D.~G. 2003, \apj,
  584, 210

\bibitem[{{Gal} {et~al.}(2008){Gal}, {Lemaux}, {Lubin}, {Kocevski}, \&
  {Squires}}]{gal:08}
{Gal}, R.~R., {Lemaux}, B.~C., {Lubin}, L.~M., {Kocevski}, D., \& {Squires},
  G.~K. 2008, \apj, 684, 933

\bibitem[{{Gallazzi} {et~al.}(2008){Gallazzi}, {Bell}, {Wolf}, {Gray},
  {Papovich}, {Barden}, {Peng}, \& et~al.}]{gallazzi:08}
{Gallazzi}, A., {Bell}, E.~F., {Wolf}, C., {Gray}, M.~E., {Papovich}, C.,
  {Barden}, M., {Peng}, C.~Y., \& et~al. 2008, ArXiv e-prints

\bibitem[{{Geller} \& {Huchra}(1983)}]{geller:83}
{Geller}, M.~J. \& {Huchra}, J.~P. 1983, \apjs, 52, 61

\bibitem[{{Gonzalez} {et~al.}(2005){Gonzalez}, {Tran}, {Conbere}, \&
  {Zaritsky}}]{gonzalez:05}
{Gonzalez}, A.~H., {Tran}, K.~H., {Conbere}, M.~N., \& {Zaritsky}, D. 2005,
  \apjl, 624, L73

\bibitem[{{Gray} {et~al.}(2002){Gray}, {Taylor}, {Meisenheimer}, {Dye}, {Wolf},
  \& {Thommes}}]{gray:02}
{Gray}, M.~E., {Taylor}, A.~N., {Meisenheimer}, K., {Dye}, S., {Wolf}, C., \&
  {Thommes}, E. 2002, \apj, 568, 141

\bibitem[{{Gregory} \& {Thompson}(1978)}]{gregory:78}
{Gregory}, S.~A. \& {Thompson}, L.~A. 1978, \apj, 222, 784

\bibitem[{{Hashimoto} {et~al.}(1998){Hashimoto}, {Oemler}, {Lin}, \&
  {Tucker}}]{hashimoto:98}
{Hashimoto}, Y., {Oemler}, A.~J., {Lin}, H., \& {Tucker}, D.~L. 1998, \apj,
  499, 589

\bibitem[{{Hester}(2006)}]{hester:06}
{Hester}, J.~A. 2006, \apj, 647, 910

\bibitem[{{Hoekstra} {et~al.}(2002){Hoekstra}, {Franx}, {Kuijken}, \& {van
  Dokkum}}]{hoekstra:02}
{Hoekstra}, H., {Franx}, M., {Kuijken}, K., \& {van Dokkum}, P.~G. 2002,
  \mnras, 333, 911+

\bibitem[{{Holden} {et~al.}(2007){Holden}, {Illingworth}, {Franx}, {Blakeslee},
  \& et~al.}]{holden:07}
{Holden}, B.~P., {Illingworth}, G.~D., {Franx}, M., {Blakeslee}, J.~P., \&
  et~al. 2007, \apj, 670, 190

\bibitem[{{Jeltema} {et~al.}(2007){Jeltema}, {Mulchaey}, {Lubin}, \&
  {Fassnacht}}]{jeltema:07}
{Jeltema}, T.~E., {Mulchaey}, J.~S., {Lubin}, L.~M., \& {Fassnacht}, C.~D.
  2007, \apj, 658, 865

\bibitem[{{Kautsch} {et~al.}(2008){Kautsch}, {Gonzalez}, {Soto}, {Tran},
  {Zaritsky}, \& {Moustakas}}]{kautsch:08}
{Kautsch}, S.~J., {Gonzalez}, A.~H., {Soto}, C.~A., {Tran}, K.-V.~H.,
  {Zaritsky}, D., \& {Moustakas}, J. 2008, \apjl, 688, L5

\bibitem[{{Kawata} \& {Mulchaey}(2008)}]{kawata:08}
{Kawata}, D. \& {Mulchaey}, J.~S. 2008, \apjl, 672, L103

\bibitem[{{Kelson} {et~al.}(2000){Kelson}, {Illingworth}, {van Dokkum}, \&
  {Franx}}]{kelson:00b}
{Kelson}, D.~D., {Illingworth}, G.~D., {van Dokkum}, P.~G., \& {Franx}, M.
  2000, \apj, 531, 159

\bibitem[{{Knobel} {et~al.}(2009){Knobel}, {Lilly}, {Iovino}, {Porciani},
  {Kova{\v c}}, {Cucciati}, {Finoguenov}, \& et~al.}]{knobel:09}
{Knobel}, C., {Lilly}, S.~J., {Iovino}, A., {Porciani}, C., {Kova{\v c}}, K.,
  {Cucciati}, O., {Finoguenov}, A., \& et~al. 2009, \apj, 697, 1842

\bibitem[{{Kocevski} {et~al.}(2009){Kocevski}, {Lubin}, {Gal}, {Lemaux},
  {Fassnacht}, \& {Squires}}]{kocevski:09}
{Kocevski}, D.~D., {Lubin}, L.~M., {Gal}, R., {Lemaux}, B.~C., {Fassnacht},
  C.~D., \& {Squires}, G.~K. 2009, \apj, 690, 295

\bibitem[{{Koyama} {et~al.}(2008){Koyama}, {Kodama}, {Shimasaku}, {Okamura}, \&
  et~al.}]{koyama:08}
{Koyama}, Y., {Kodama}, T., {Shimasaku}, K., {Okamura}, S., \& et~al. 2008,
  \mnras, 391, 1758

\bibitem[{{Le Floc'h} {et~al.}(2005){Le Floc'h}, {Papovich}, {Dole}, {Bell},
  {Lagache}, {Rieke}, {Egami}, {P{\'e}rez-Gonz{\'a}lez}, \&
  et~al.}]{lefloch:05}
{Le Floc'h}, E., {Papovich}, C., {Dole}, H., {Bell}, E.~F., {Lagache}, G.,
  {Rieke}, G.~H., {Egami}, E., {P{\'e}rez-Gonz{\'a}lez}, P.~G., \& et~al. 2005,
  \apj, 632, 169

\bibitem[{{LeFevre} {et~al.}(2003){LeFevre}, {Saisse}, {Mancini}, {Brau-Nogue},
  \& et~al.}]{lefevre:03}
{LeFevre}, O., {Saisse}, M., {Mancini}, D., {Brau-Nogue}, S., \& et~al. 2003,
  in Presented at the Society of Photo-Optical Instrumentation Engineers (SPIE)
  Conference, Vol. 4841, Society of Photo-Optical Instrumentation Engineers
  (SPIE) Conference Series, ed. M.~{Iye} \& A.~F.~M. {Moorwood}, 1670--1681

\bibitem[{{Lidman} {et~al.}(2008){Lidman}, {Rosati}, {Tanaka}, {Strazzullo},
  {Demarco}, {Mullis}, {Ageorges}, {Kissler-Patig}, {Petr-Gotzens}, \&
  {Selman}}]{lidman:08}
{Lidman}, C., {Rosati}, P., {Tanaka}, M., {Strazzullo}, V., {Demarco}, R.,
  {Mullis}, C., {Ageorges}, N., {Kissler-Patig}, M., {Petr-Gotzens}, M.~G., \&
  {Selman}, F. 2008, \aap, 489, 981

\bibitem[{{Lotz} {et~al.}(2008){Lotz}, {Davis}, {Faber}, {Guhathakurta},
  {Gwyn}, {Huang}, {Koo}, {Le Floc'h}, \& et~al.}]{lotz:08}
{Lotz}, J.~M., {Davis}, M., {Faber}, S.~M., {Guhathakurta}, P., {Gwyn}, S.,
  {Huang}, J., {Koo}, D.~C., {Le Floc'h}, E., \& et~al. 2008, \apj, 672, 177

\bibitem[{{Martini} {et~al.}(2007){Martini}, {Mulchaey}, \&
  {Kelson}}]{martini:07}
{Martini}, P., {Mulchaey}, J.~S., \& {Kelson}, D.~D. 2007, \apj, 664, 761

\bibitem[{{Marzke} {et~al.}(1998){Marzke}, {da Costa}, {Pellegrini}, {Willmer},
  \& {Geller}}]{marzke:98}
{Marzke}, R.~O., {da Costa}, L.~N., {Pellegrini}, P.~S., {Willmer}, C.~N.~A.,
  \& {Geller}, M.~J. 1998, \apj, 503, 617

\bibitem[{{McCarthy} {et~al.}(2008){McCarthy}, {Frenk}, {Font}, {Lacey},
  {Bower}, {Mitchell}, {Balogh}, \& {Theuns}}]{mccarthy:08}
{McCarthy}, I.~G., {Frenk}, C.~S., {Font}, A.~S., {Lacey}, C.~G., {Bower},
  R.~G., {Mitchell}, N.~L., {Balogh}, M.~L., \& {Theuns}, T. 2008, \mnras, 383,
  593

\bibitem[{{Mei} {et~al.}(2009){Mei}, {Holden}, {Blakeslee}, {Ford}, {Franx},
  {Homeier}, {Illingworth}, \& et~al.}]{mei:09}
{Mei}, S., {Holden}, B.~P., {Blakeslee}, J.~P., {Ford}, H.~C., {Franx}, M.,
  {Homeier}, N.~L., {Illingworth}, G.~D., \& et~al. 2009, \apj, 690, 42

\bibitem[{{Moore} {et~al.}(1998){Moore}, {Lake}, \& {Katz}}]{moore:98}
{Moore}, B., {Lake}, G., \& {Katz}, N. 1998, \apj, 495, 139+

\bibitem[{{Moran} {et~al.}(2007){Moran}, {Loh}, {Ellis}, {Treu}, {Bundy}, \&
  {MacArthur}}]{moran:07}
{Moran}, S.~M., {Loh}, B.~L., {Ellis}, R.~S., {Treu}, T., {Bundy}, K., \&
  {MacArthur}, L.~A. 2007, \apj, 665, 1067

\bibitem[{{Moustakas} {et~al.}(2006){Moustakas}, {Kennicutt}, \&
  {Tremonti}}]{moustakas:06}
{Moustakas}, J., {Kennicutt}, Jr., R.~C., \& {Tremonti}, C.~A. 2006, \apj, 642,
  775

\bibitem[{{Muzzin} {et~al.}(2008){Muzzin}, {Wilson}, {Lacy}, {Yee}, \&
  {Stanford}}]{muzzin:08}
{Muzzin}, A., {Wilson}, G., {Lacy}, M., {Yee}, H.~K.~C., \& {Stanford}, S.~A.
  2008, \apj, 686, 966

\bibitem[{{O'Donnell}(1994)}]{odonnell:94}
{O'Donnell}, J.~E. 1994, \apj, 422, 158

\bibitem[{Oke {et~al.}(1995)Oke, Cohen, Carr, Cromer, Dingizian, Harris,
  Labrecque, Luciano, Schaal, Epps, \& Miller}]{oke:95}
Oke, J.~B., Cohen, J.~G., Carr, M., Cromer, J., Dingizian, A., Harris, F.~H.,
  Labrecque, S., Luciano, R., Schaal, W., Epps, H., \& Miller, J. 1995, \pasp,
  107, 375

\bibitem[{{Peebles}(1970)}]{peebles:70}
{Peebles}, P.~J.~E. 1970, \aj, 75, 13

\bibitem[{{Poggianti} {et~al.}(2009){Poggianti}, {Arag{\'o}n-Salamanca},
  {Zaritsky}, {DeLucia}, {Milvang-Jensen}, {Desai}, {Jablonka}, \&
  et~al.}]{poggianti:09}
{Poggianti}, B.~M., {Arag{\'o}n-Salamanca}, A., {Zaritsky}, D., {DeLucia}, G.,
  {Milvang-Jensen}, B., {Desai}, V., {Jablonka}, P., \& et~al. 2009, \apj, 693,
  112

\bibitem[{{Poggianti} {et~al.}(2001){Poggianti}, {Bridges}, {Carter},
  {Mobasher}, {Doi}, {Iye}, {Kashikawa}, {Komiyama}, {Okamura}, {Sekiguchi},
  {Shimasaku}, {Yagi}, \& {Yasuda}}]{poggianti:01b}
{Poggianti}, B.~M., {Bridges}, T.~J., {Carter}, D., {Mobasher}, B., {Doi}, M.,
  {Iye}, M., {Kashikawa}, N., {Komiyama}, Y., {Okamura}, S., {Sekiguchi}, M.,
  {Shimasaku}, K., {Yagi}, M., \& {Yasuda}, N. 2001, \apj, 563, 118

\bibitem[{{Poggianti} {et~al.}(2004){Poggianti}, {Bridges}, {Komiyama}, {Yagi},
  {Carter}, {Mobasher}, {Okamura}, \& {Kashikawa}}]{poggianti:04}
{Poggianti}, B.~M., {Bridges}, T.~J., {Komiyama}, Y., {Yagi}, M., {Carter}, D.,
  {Mobasher}, B., {Okamura}, S., \& {Kashikawa}, N. 2004, \apj, 601, 197

\bibitem[{{Poggianti} {et~al.}(2008){Poggianti}, {Desai}, {Finn}, {Bamford},
  {De Lucia}, {Varela}, {Arag{\'o}n-Salamanca}, \& et~al.}]{poggianti:08}
{Poggianti}, B.~M., {Desai}, V., {Finn}, R., {Bamford}, S., {De Lucia}, G.,
  {Varela}, J., {Arag{\'o}n-Salamanca}, A., \& et~al. 2008, \apj, 684, 888

\bibitem[{{Postman} {et~al.}(2005){Postman}, {Franx}, {Cross}, {Holden},
  {Ford}, {Illingworth}, \& {et al.}}]{postman:05}
{Postman}, M., {Franx}, M., {Cross}, N.~J.~G., {Holden}, B., {Ford}, H.~C.,
  {Illingworth}, G.~D., \& {et al.} 2005, \apj, 623, 721

\bibitem[{{Rasmussen} {et~al.}(2008){Rasmussen}, {Ponman}, {Verdes-Montenegro},
  {Yun}, \& {Borthakur}}]{rasmussen:08}
{Rasmussen}, J., {Ponman}, T.~J., {Verdes-Montenegro}, L., {Yun}, M.~S., \&
  {Borthakur}, S. 2008, \mnras, 388, 1245

\bibitem[{{Rieke} {et~al.}(2009){Rieke}, {Alonso-Herrero}, {Weiner},
  {P{\'e}rez-Gonz{\'a}lez}, {Blaylock}, {Donley}, \& {Marcillac}}]{rieke:09}
{Rieke}, G.~H., {Alonso-Herrero}, A., {Weiner}, B.~J.,
  {P{\'e}rez-Gonz{\'a}lez}, P.~G., {Blaylock}, M., {Donley}, J.~L., \&
  {Marcillac}, D. 2009, \apj, 692, 556

\bibitem[{{Rieke} {et~al.}(2004){Rieke}, {Young}, {Engelbracht}, {Kelly},
  {Low}, {Haller}, {Beeman}, \& et~al.}]{rieke:04}
{Rieke}, G.~H., {Young}, E.~T., {Engelbracht}, C.~W., {Kelly}, D.~M., {Low},
  F.~J., {Haller}, E.~E., {Beeman}, J.~W., \& et~al. 2004, \apjs, 154, 25

\bibitem[{{Romeo} {et~al.}(2008){Romeo}, {Napolitano}, {Covone},
  {Sommer-Larsen}, {Antonuccio-Delogu}, \& {Capaccioli}}]{romeo:08}
{Romeo}, A.~D., {Napolitano}, N.~R., {Covone}, G., {Sommer-Larsen}, J.,
  {Antonuccio-Delogu}, V., \& {Capaccioli}, M. 2008, \mnras, 389, 13

\bibitem[{{Saintonge} {et~al.}(2008){Saintonge}, {Tran}, \&
  {Holden}}]{saintonge:08}
{Saintonge}, A., {Tran}, K.-V.~H., \& {Holden}, B.~P. 2008, \apjl, 685, L113

\bibitem[{{Salpeter}(1955)}]{salpeter:55}
{Salpeter}, E.~E. 1955, \apj, 121, 161

\bibitem[{{Schechter}(1976)}]{schechter:76}
{Schechter}, P. 1976, \apj, 203, 297

\bibitem[{{Schlegel} {et~al.}(1998){Schlegel}, {Finkbeiner}, \&
  {Davis}}]{schlegel:98}
{Schlegel}, D.~J., {Finkbeiner}, D.~P., \& {Davis}, M. 1998, \apj, 500, 525+

\bibitem[{{Stott} {et~al.}(2007){Stott}, {Smail}, {Edge}, {Ebeling}, {Smith},
  {Kneib}, \& {Pimbblet}}]{stott:07}
{Stott}, J.~P., {Smail}, I., {Edge}, A.~C., {Ebeling}, H., {Smith}, G.~P.,
  {Kneib}, J.-P., \& {Pimbblet}, K.~A. 2007, \apj, 661, 95

\bibitem[{{Tanaka} {et~al.}(2007){Tanaka}, {Kodama}, {Kajisawa}, {Bower},
  {Demarco}, {Finoguenov}, {Lidman}, \& {Rosati}}]{tanaka:07}
{Tanaka}, M., {Kodama}, T., {Kajisawa}, M., {Bower}, R., {Demarco}, R.,
  {Finoguenov}, A., {Lidman}, C., \& {Rosati}, P. 2007, \mnras, 377, 1206

\bibitem[{{Tonnesen} {et~al.}(2007){Tonnesen}, {Bryan}, \& {van
  Gorkom}}]{tonnesen:07}
{Tonnesen}, S., {Bryan}, G.~L., \& {van Gorkom}, J.~H. 2007, \apj, 671, 1434

\bibitem[{{Tran} {et~al.}(2003){Tran}, {Franx}, {Illingworth}, {Kelson}, \&
  {van Dokkum}}]{tran:03b}
{Tran}, K.~H., {Franx}, M., {Illingworth}, G., {Kelson}, D.~D., \& {van
  Dokkum}, P. 2003, \apj, 599, 865

\bibitem[{{Tran} {et~al.}(2004){Tran}, {Franx}, {Illingworth}, {van Dokkum},
  {Kelson}, \& {Magee}}]{tran:04a}
{Tran}, K.~H., {Franx}, M., {Illingworth}, G.~D., {van Dokkum}, P., {Kelson},
  D.~D., \& {Magee}, D. 2004, \apj, 609, 683

\bibitem[{{Tran} {et~al.}(2001){Tran}, {Simard}, {Zabludoff}, \&
  {Mulchaey}}]{tran:01}
{Tran}, K.~H., {Simard}, L., {Zabludoff}, A.~I., \& {Mulchaey}, J.~S. 2001,
  \apj, 549, 172

\bibitem[{{Tran} {et~al.}(2005){Tran}, {van Dokkum}, {Illingworth}, {Kelson},
  {Gonzalez}, \& {Franx}}]{tran:05a}
{Tran}, K.~H., {van Dokkum}, P., {Illingworth}, G.~D., {Kelson}, D.,
  {Gonzalez}, A., \& {Franx}, M. 2005, \apj, 619, 134

\bibitem[{{Tran} {et~al.}(2007){Tran}, {Franx}, {Illingworth}, {van Dokkum},
  {Kelson}, {Blakeslee}, \& {Postman}}]{tran:07}
{Tran}, K.-V.~H., {Franx}, M., {Illingworth}, G.~D., {van Dokkum}, P.,
  {Kelson}, D.~D., {Blakeslee}, J.~P., \& {Postman}, M. 2007, \apj, 661, 750

\bibitem[{{Tran} {et~al.}(2008){Tran}, {Moustakas}, {Gonzalez}, {Bai},
  {Zaritsky}, \& {Kautsch}}]{tran:08}
{Tran}, K.-V.~H., {Moustakas}, J., {Gonzalez}, A.~H., {Bai}, L., {Zaritsky},
  D., \& {Kautsch}, S.~J. 2008, \apjl, 683, L17

\bibitem[{{van Dokkum} {et~al.}(2000){van Dokkum}, {Franx}, {Fabricant},
  {Illingworth}, \& {Kelson}}]{vandokkum:00}
{van Dokkum}, P.~G., {Franx}, M., {Fabricant}, D., {Illingworth}, G.~D., \&
  {Kelson}, D.~D. 2000, \apj, 541, 95

\bibitem[{{van Dokkum} {et~al.}(1998{\natexlab{a}}){van Dokkum}, {Franx},
  {Kelson}, \& {Illingworth}}]{vandokkum:98b}
{van Dokkum}, P.~G., {Franx}, M., {Kelson}, D.~D., \& {Illingworth}, G.~D.
  1998{\natexlab{a}}, \apjl, 504, L17

\bibitem[{{van Dokkum} {et~al.}(1998{\natexlab{b}}){van Dokkum}, {Franx},
  {Kelson}, {Illingworth}, {Fisher}, \& {Fabricant}}]{vandokkum:98a}
{van Dokkum}, P.~G., {Franx}, M., {Kelson}, D.~D., {Illingworth}, G.~D.,
  {Fisher}, D., \& {Fabricant}, D. 1998{\natexlab{b}}, \apj, 500, 714+

\bibitem[{{Wilman} {et~al.}(2008){Wilman}, {Pierini}, {Tyler}, {McGee},
  {Oemler}, {Morris}, {Balogh}, {Bower}, \& {Mulchaey}}]{wilman:08}
{Wilman}, D.~J., {Pierini}, D., {Tyler}, K., {McGee}, S.~L., {Oemler}, Jr., A.,
  {Morris}, S.~L., {Balogh}, M.~L., {Bower}, R.~G., \& {Mulchaey}, J.~S. 2008,
  \apj, 680, 1009

\bibitem[{{Wolf} {et~al.}(2009){Wolf}, {Arag{\'o}n-Salamanca}, {Balogh},
  {Barden}, {Bell}, {Gray}, {Peng}, \& et~al.}]{wolf:09}
{Wolf}, C., {Arag{\'o}n-Salamanca}, A., {Balogh}, M., {Barden}, M., {Bell},
  E.~F., {Gray}, M.~E., {Peng}, C.~Y., \& et~al. 2009, \mnras, 393, 1302

\bibitem[{{Yang} {et~al.}(2007){Yang}, {Mo}, {van den Bosch}, {Pasquali}, {Li},
  \& {Barden}}]{yang:07}
{Yang}, X., {Mo}, H.~J., {van den Bosch}, F.~C., {Pasquali}, A., {Li}, C., \&
  {Barden}, M. 2007, \apj, 671, 153

\bibitem[{{Yee} {et~al.}(1996){Yee}, {Ellingson}, \& {Carlberg}}]{yee:96}
{Yee}, H.~K.~C., {Ellingson}, E., \& {Carlberg}, R.~G. 1996, \apjs, 102, 269+

\bibitem[{{Zabludoff} \& {Mulchaey}(1998)}]{zabludoff:98a}
{Zabludoff}, A.~I. \& {Mulchaey}, J.~S. 1998, \apj, 496, 39

\end{thebibliography}

\clearpage
\begin{deluxetable}{rrrrrr}
\tablecaption{Properties of Galaxy Groups In SG1120\label{tab:groups}}
\tablewidth{0pt}
\tablehead{
\colhead{ID}   & 
\colhead{$(\alpha,\delta)$\tablenotemark{a}} &
\colhead{$\bar{z}$~\tablenotemark{b}}   &
\colhead{$\sigma_{1D}$\tablenotemark{b}}   & 
\colhead{$N_{g}$\tablenotemark{b}} &
\colhead{$T_X$\tablenotemark{c}} 
}
\startdata
G1 & (11:20:07.48, -12:05:09.1) & $0.3522\pm0.0008$ & $303\pm60$ & 13 & 2.2\\
G2 & (11:20:13.33, -11:58:50.6) & $0.3707\pm0.0007$ & $406\pm83$ & 21 & 1.7\\
G3 & (11:20:22.19, -12:01:46.1) & $0.3712\pm0.0005$ & $580\pm100$ & 35 & 1.8\\
G4 & (11:20:10.14, -12:08:51.6) & $0.3694\pm0.0005$ & $567\pm119$ & 22 & 3.0\\
\enddata
\tablenotetext{a}{Coordinates (J2000) of the brightest galaxy in each
group.}  
\tablenotetext{b}{Mean redshift ($\bar{z}$) and line-of-sight velocity
dispersion ($\sigma_{1D}$; km~s$^{-1}$) determined using galaxies
($N_g$) within 500~kpc of the brightest group galaxy; $\bar{z}$ and
$\sigma_{1D}$ are determined using the biweight and jackknife
estimators \citep{beers:90}, respectively. } 
\tablenotetext{c}{X-ray temperatures (keV) from \citet{gonzalez:05}.}
\end{deluxetable}

\begin{deluxetable}{llrrr}
\tablewidth{0pc}
\tablecaption{MIPS \mipsmu~Galaxy Fractions\label{tab:fractions}}
\tablehead{
\colhead{Selection} & \colhead{Number} &  
\colhead{Field} & \colhead{Group} & \colhead{Cluster}}
\startdata
All\tablenotemark{a}	& $N$	   & 87 & 143 & 171 \\
\cutinhead{Luminosity-Selected Samples}
$M_V<$\magcut 		& $N$ 	   & 29 &  98 & 105 \\
			& $N_{IR}$ & 11 &  31 & 7 \\
			& IR\% 	   & 37.9\% & 31.6\% & 6.7\% \\
$M_V<$\magcut			& & & & \\
~~~Late-types\tablenotemark{b}  & $N$	   & 18 & 37 & 21 \\ 
				& $N_{IR}$ & 11 & 27 & 7 \\
				& IR\% & 61.1\% & 73.0\% & 33.3\% \\
$M_V<$\magcut			& & & & \\
~~~Early-types\tablenotemark{b} & $N$	   & 8  & 61 & 80 \\
				& $N_{IR}$ & 0  &  4 & 0 \\
				& IR\%	& 0.0\% & 6.6\% & 0.0\% \\ 
\cutinhead{Mass-Selected Samples}
$\log(M_{\ast})[M_{\odot}]>10.6$ & $N$	   & 22 & 72 & 103 \\
				& $N_{IR}$ & 8  & 14  & 5 \\
				& IR\%	   & 36.4\% & 19.4\% & 4.9\% \\ 
$\log(M_{\ast})[M_{\odot}]>10.6$ & & & & \\
~~~Late-types\tablenotemark{b}  & $N$	   & 12 & 16 & 18 \\ 
				& $N_{IR}$ & 8  & 11 & 5 \\
				& IR\% & 66.7\% & 68.8\% & 27.8\% \\ 
$\log(M_{\ast})[M_{\odot}]>10.6$ & & & & \\
~~~Early-types\tablenotemark{b} & $N$	   & 8  & 56 & 81 \\
				& $N_{IR}$ & 0  &  3 & 0 \\
				& IR\%	& 0.0\% & 5.3\% & 0.0\% \\
\enddata
\tablenotetext{a}{Considering only spectroscopically confirmed members
that fall on the HST imaging.}
\tablenotetext{b}{{Late-type (disk-dominated) galaxies have Hubble
classification of $T>0$ and early-type (bulge-dominated) galaxies have
$T\leq0$.}}
\end{deluxetable}

\begin{deluxetable}{llrr}
\tablecaption{Infrared Luminosity Function Parameters\label{tab:IRLF}}
\tablewidth{0pt}
\tablehead{
\colhead{Function} & \colhead{Environment} &  
\colhead{$\log(L^{\ast}_{IR})$} & \colhead{$\phi^{\ast}_{IR}$}	}
\startdata
Schechter\tablenotemark{a} & Group	& $44.99^{+0.19}_{-0.19}$ & $3.4^{+1.4}_{-0.9}$ \\
Schechter\tablenotemark{a} & Cluster	& $44.33^{+0.32}_{-0.25}$ & $5.3^{+4.0}_{-3.5}$ \\
Schechter\tablenotemark{a} & Evolved\tablenotemark{b}	& 44.53 & 4.0 \\ \\
Double-exponential\tablenotemark{c} & Group	& $43.71^{+0.19}_{-0.19}$ & $11.8^{+4.7}_{-3.2}$ \\
Double-exponential\tablenotemark{c} & Cluster	& $43.08^{+0.37}_{0.46}$ & $17.4^{+32.2}_{-8.8}$ \\
Double-exponential\tablenotemark{c} & (Field)\tablenotemark{d} & $43.28^{+0.09}_{0.03}$ & \nodata \\
\enddata
\tablenotetext{a}{For the Schechter profile, we set
$\alpha_{IR}=1.414$ \citep{bai:06,bai:07}.}
\tablenotetext{b}{IR LF measured in $z\sim0$ galaxy clusters
\citep{bai:06} evolved to $z\sim0.35$ using the evolution measured in
the field IR LF \citep{lefloch:05}.} 
\tablenotetext{c}{For the double-exponential profile, we set
$\alpha_{IR}=1.23$ and $\sigma_{IR}=0.72$ \citep{lefloch:05}}
\tablenotetext{d}{Field $L^{\ast}_{IR}$ measured by
\citet{lefloch:05} for galaxies at $(0.3<z<0.45)$; we do not include 
$\phi_{IR}^{\ast}$ because it is normalized differently in the field
and in clusters.} 
\end{deluxetable}

\clearpage

\clearpage

\end{document}